\pdfoutput=1

\documentclass{jfm}

\usepackage{graphicx,tikz,array,amsmath}
\usepackage{natbib}

\ifCUPmtlplainloaded \else
  \checkfont{eurm10}
  \iffontfound
    \IfFileExists{upmath.sty}
      {\typeout{^^JFound AMS Euler Roman fonts on the system,
                   using the 'upmath' package.^^J}%
       \usepackage{upmath}}
      {\typeout{^^JFound AMS Euler Roman fonts on the system, but you
                   dont seem to have the}%
       \typeout{'upmath' package installed. JFM.cls can take advantage
                 of these fonts,^^Jif you use 'upmath' package.^^J}%
      }
  \else
  \fi
\fi

\ifCUPmtlplainloaded \else
  \checkfont{msam10}
  \iffontfound
    \IfFileExists{amssymb.sty}
      {\typeout{^^JFound AMS Symbol fonts on the system, using the
                'amssymb' package.^^J}%
       \usepackage{amssymb}%
       \let\le=\leqslant  
       \let\ge=\geqslant  
      }{}
  \fi
\fi

\ifCUPmtlplainloaded \else
  \IfFileExists{amsbsy.sty}
    {\typeout{^^JFound the 'amsbsy' package on the system, using it.^^J}%
     \usepackage{amsbsy}}
    {\providecommand\boldsymbol[1]{\mbox{\boldmath $##1$}}}
\fi

\graphicspath{{figs/}{frames/}}

\newcommand\solidrule[1][21pt]{\rule[0.5ex]{#1}{.4pt}}
\newcommand\dashedrule{\mbox{%
	\solidrule[5pt]\hspace{3pt}\solidrule[5pt]\hspace{3pt}\solidrule[5pt]}}

\newcommand\dashdotrule{\mbox{%
	\solidrule[4pt]\hspace{2pt}\solidrule[1pt]\hspace{2pt}\solidrule[4pt]\hspace{2pt}\solidrule[1pt]%
	\hspace{2pt}\solidrule[4pt]}}
	
\newcommand{\bu}{\mathbf u}
\newcommand{\oT}{\overline T}
\newcommand{\wT}{\langle wT\rangle}

\newcommand{\e}{\cdot10^}

\newcommand\Pran{Pr} 

\newsavebox{\astrutbox}
\sbox{\astrutbox}{\rule[-5pt]{0pt}{20pt}}

\newcommand\etal{\mbox{\textit{et al.}}}

\title[Convectively driven shear and decreased heat flux]{Convectively driven shear and \\decreased heat flux}

\author[D.\ Goluskin \etal]
{David Goluskin$^1$%
\thanks{Email address for correspondence: goluskin@umich.edu}
\thanks{Present address: Mathematics Department, University of Michigan, Ann Arbor, MI, USA}
,\ns Hans Johnston$^2$,\ns Glenn R.\ Flierl$^3$,\\\ns Edward A.\ Spiegel$^{4,5}$}
\affiliation{$^1$Department of Applied Physics and Applied Mathematics, Columbia University\\
New York, NY, USA\\
[\affilskip]$^2$Department of Mathematics and Statistics, University of Massachusetts\\
Amherst, MA, USA\\
[\affilskip]$^3$Department of Earth, Atmospheric, and Planetary Sciences\\Massachusetts Institute of Technology,
Cambridge, MA, USA\\
[\affilskip]$^4$Department of Astronomy, Columbia University, New York, NY, USA\\
[\affilskip]$^5$Visiting Scholar, New York University, New York, NY, USA}

\pubyear{2013}
\volume{?}
\pagerange{?}
\date{?; revised ?; accepted ?. - To be entered by editorial office}

\begin{document}

\maketitle

\begin{abstract}
We report on direct numerical simulations of two-dimensional, horizontally periodic Rayleigh-B\'enard convection, focusing on its ability to drive large-scale horizontal flow that is vertically sheared. For the Prandtl numbers ($\Pran$) between 1 and 10 simulated here, this large-scale shear can be induced by raising the Rayleigh number ($Ra$) sufficiently, and we explore the resulting convection for $Ra$ up to $10^{10}$. When present in our simulations, the sheared mean flow accounts for a large fraction of the total kinetic energy, and this fraction tends towards unity as $Ra\to\infty$. The shear helps disperse convective structures, and it reduces vertical heat flux; in parameter regimes where one state with large-scale shear and one without are both stable, the Nusselt number of the state with shear is smaller and grows more slowly with $Ra$. When the large-scale shear is present with $\Pran\lesssim2$, the convection undergoes strong global oscillations on long timescales, and heat transport occurs in bursts. Nusselt numbers, time-averaged over these bursts, vary non-monotonically with $Ra$ for $\Pran=1$. When the shear is present with $\Pran\gtrsim3$, the flow does not burst, and convective heat transport is sustained at all times. Nusselt numbers then grow roughly as powers of $Ra$, but the growth rates are slower than any previously reported for Rayleigh-B\'enard convection without large-scale shear. We find the Nusselt numbers grow proportionally to $Ra^{0.077}$ when $\Pran=3$ and to $Ra^{0.19}$ when $\Pran=10$. Analogies with tokamak plasmas are described.
\end{abstract}

\begin{keywords}
\end{keywords}

\section{Introduction}
\label{sec: intro}

Though buoyancy forces in Rayleigh-B\'enard (RB) convection act only vertically, pressure drives horizontal motion as well. For some geometries and boundary conditions, mean horizontal motion can arise that is stronger than the vertical motion by several orders of magnitude. Here we simulate RB convection in a two-dimensional layer with periodicity imposed in the horizontal direction, focusing on states with horizontal mean flow at moderate-to-large $Ra$. Such mean flows are vertically sheared, and this large-scale shear alters many bulk properties of the convection by shearing all smaller flow structures in the same way. Hence, we refer to such states as \emph{shearing convection}. States without mean shear, which might be thought of as ordinary RB convection, we refer to as \emph{non-shearing convection}.  By analogy with wavenumber-zero mean flows in planetary atmospheres and tokamak plasmas \citep{Diamond2005}, which are discussed below, we refer to the mean horizontal flow as \emph{zonal flow}. Both zonal flow and non-zonal flow are present in shearing convection, whereas the zonal component either is absent from non-shearing convection or vanishes in long-time averages.

\begin{figure}
\centering
\begin{tikzpicture}
\node at (0,1.25) {Cold, free-slip};
\node at (0,-1.25) {Hot, free-slip};
\draw[black,thick] (-1.8,1) -- (1.8,1);
\draw[black,thick] (-1.8,-1) -- (1.8,-1);
\draw[black] (-.9,.7) arc (90:270:.7);
\draw[black,-latex] (-.9,-.7) arc (-90:70:.7);
\draw[black] (.9,-.7) arc (-90:90:.7);
\draw[black,-latex] (.9,-.7) arc (270:110:.7);
\draw[black,-latex,thick] (1.9,.3) -- (1.9,-.3);
\node at (2.1,.1) {$\mathbf{g}$};
\node at (-2.15,1.25) {(a)};
\draw[black,thick] (3.2,1) -- (6.8,1);
\draw[black,thick] (3.2,-1) -- (6.8,-1);
\draw[black,densely dotted] (5,-1) -- (5,1); 
\draw[black] (5,0) to [out=225,in=90] (4.5,-1);
\draw[black] (5,0) to [out=45,in=270] (5.5,1);
\draw[black,-stealth] (5,.8) -- (5.48,.8);
\draw[black,-stealth] (5,.6) -- (5.43,.6);
\draw[black,-stealth] (5,.4) -- (5.33,.4);
\draw[black,-stealth] (5,.2) -- (5.185,.2);
\draw[black,-stealth] (5,-.8) -- (4.52,-.8);
\draw[black,-stealth] (5,-.6) -- (4.57,-.6);
\draw[black,-stealth] (5,-.4) -- (4.67,-.4);
\draw[black,-stealth] (5,-.2) -- (4.815,-.2);
\node at (2.85,1.25) {(b)};
\end{tikzpicture}
\caption{The configuration of 2D RB convection studied here with schematic drawings of (a) the vortical motions typical of 2D convection and (b) the sort of mean zonal flow that can develop with free-slip boundary conditions. The downward direction of gravity ($\mathbf{g}$) is indicated.}
\label{fig: RB}
\end{figure}
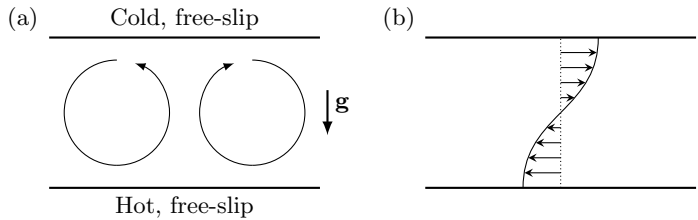

Our configuration is shown in figure \ref{fig: RB}, along with schematic drawings of the vortical motions that typically drive 2D convection (left) and the sort of mean zonal flow profile that can arise (right). Free-slip conditions are imposed at the top and bottom boundaries. This configuration, perhaps more than any other, encourages very strong mean flows: horizontal periodicity allows for mean flow with a horizontal wavenumber of zero, free-slip boundaries apply no shear stresses to slow the fluid, and two-dimensionality precludes transverse perturbations that can disrupt the mean flow. The zonal flow will always be vertically sheared since conservation of horizontal momentum requires any mean flow developing from static fluid to consist of both left-going and right-going regions, as in figure \hbox{\ref{fig: RB}(b)}. Our chosen geometry is the simplest finite domain in which convection can drive wavenumber-zero mean flow. Other examples, all of which must have a periodic coordinate, include azimuthal mean flow in an annulus and longitudinal mean flow in a spherical shell.

The shearing convection we study here differs in several ways from the ordinary (non-shearing) RB convection described, for example, in the review of \citet{Ahlers2009}. Because zonal flow helps disperse thermal plumes \citep{Terry2000}, it can significantly depress the net heat transport, along with the accompanying work expended by buoyancy, thereby inhibiting its own source of energy. These trends are qualitatively understood, but quantitative questions remain. In particular, for what parameters can shearing convection persist? And, when it does, how energetic will the zonal flow be, and how quickly will heat be transported?

Past studies of shearing convection have been motivated either by experimental or planetary convection or by plasmas in tokamaks. These applications share several features with our RB model. First, at least one boundary is often modelled as free-slip in the study of both atmospheres and tokamaks. Second, the zonal flows that form are sheared normally to their directions of motion and inhibit transport in this normal direction. Finally, anisotropies render fluid motion roughly two-dimensional locally. This is achieved in tokamaks by fast equilibration along magnetic field lines \citep{Wesson2011} and in rotating atmospheres by the Taylor-Proudman effect.

In horizontally isotropic 3D domains, persistent shearing convection has not been reported, nor have we found it in our own preliminary simulations. It thus appears that simulating shearing convection requires adding an anisotropic effect to 3D convection or, as in this work, imposing two-dimensionality. [The need for anisotropy in 3D is supported by the asymptotic analysis of \citet{Massaguer1992} and by the transient nature of zonal flow in the reduced model of \citet{Matthews1996}.] Zonal flow can arise also in systems driven by means other than buoyancy \citep{Drake1992, Finn1992, Childress2000}, but only with buoyancy can the zonal flow inhibit the very convective structures that drive it---an important feedback found in applications like planetary atmospheres and tokamaks.

In laboratory convection experiments, large-scale mean flows in domains of large horizontal extent have been described by \citet{Krishnamurti1981} and \citet{Malkus1954}. In rectangular and cylindrical domains with horizontal extents comparable to their heights, weaker mean ``winds'' that may drift or reverse direction are often seen in physical and numerical experiments \citep{Ahlers2009}. It is possible that finite-wavenumber circulation will resemble zonal flow more and more as the domain is widened. If so, periodic domains offer a way of studying this limit without the computational expense of simulating very wide domains. In either case, the zonal flow we study here is distinguished from the better known winds by its greater strength, its permanence once established, and its ability to significantly depress vertical heat flux.

Planetary atmospheres derive much of their kinetic energy from buoyancy forces, and it has been suggested that atmospheric zonal jets are driven primarily by the same sort of mechanism that drives zonal flow in RB convection \citep{Thompson1970, Busse1994}, wherein energy is transferred upscale by the velocity nonlinearity. In the presence of rotation, zonal flow has been seen in simulations of both 2D and 3D periodic domains \citep{Brummell1993, Christensen2002, Morin2004, Tao2010}. However, zonal jets can also be created by the redistribution of angular momentum on a rotating sphere \citep{Polvani1996}, and simulations of rotating spherical shells suggest that both mechanisms are important \citep{Heimpel2007, Kaspi2009}. The depression of heat flux by shear that we describe here also resembles the depression of transport in the sun proposed by \citet{Chaboyer1992}. Such applications invite the inclusion of rotation, which we do not confront in this work.

In toroidal tokamak devices, zonal flows in the poloidal direction are crucial to the effort to confine plasmas magnetically. This has motivated studies of models similar to the present one, as reviewed by \citet{Garcia2006}, even though some important effects are lost in 2D. If a tokamak plasma is regarded as a magnetohydrodynamic fluid, the radial ``interchange motions'' driven by centrifugal forces are analogous to RB convective motions. Interchange motions harm plasma confinement by adding to radial transport, but they can also drive poloidal zonal flow that lessens this harm, much as RB convection can drive horizontal zonal flow that reduces vertical heat flux. In plasma terminology, the onset of zonal flow is credited with the transition from the low-confinement mode to the high-confinement mode that has been seen in many devices and is considered essential for achieving fusion \citep{Terry2000, Diamond2005, Wagner2007}. Here, we use the Nusselt number to quantify the depression of heat transport by zonal flow. Plasma physicists might prefer to think of this depression as an increase in confinement.

Among past work on horizontally periodic 2D convection, shearing convection is absent from most analytical studies, perhaps because it does not occur in the weakly nonlinear regime for generic parameter values \citep{Howard1986, Rucklidge1996, GoluskinThesis}. Instead, shearing convection has been found using direct numerical simulation \citep{Thompson1970, Finn1993a, Garcia2003, Garcia2003a, Zhu2012, Bian2003a, VanderPoel2014a}, numerical bifurcation analysis \citep{Rucklidge1996}, and reduced models \citep[beginning with][]{Howard1986}. Still, parameter space has scarcely been explored at large Rayleigh numbers, despite the fact that plasma physical and astrophysical applications fall in this regime. The tokamak-motivated simulations by Garcia, Bian, and collaborators give some tantalizing first results: when zonal flow is present at a Prandtl number of unity, convective transport occurs in bursts and is, on average, significantly lessened. We expand on these findings here, computing integral quantities in a larger spatial domain and over greater ranges of the control parameters.

The governing equations are introduced in \S\ref{sec: equations}. The shearing convection we simulate can be roughly divided into two categories. At larger Prandtl numbers, the convective transport is strong at all times, so we say that the shearing convection is \emph{sustained}. At smaller Prandtl numbers, convective transport comes in pronounced bursts, so we say that the shearing convection is \emph{bursting}. Sustained and bursting flows are described in \S\ref{sec: sustained} and \S\ref{sec: bursting}, respectively. Heat transport in both sorts of shearing convection is summarized in \S\ref{sec: nusselt}. Discussion is offered throughout, and concluding remarks appear in~\S\ref{sec: con}.

\section{Governing equations}
\label{sec: equations}

We adopt the Boussinesq approximation, in which the fluid has constant kinematic viscosity, $\nu$, and thermal diffusivity, $\kappa$. We nondimensionalize lengths by the layer height, $d$, times by the thermal diffusion timescale, $d^2/\kappa$, and temperatures by the decrease, $\Delta$, from the bottom boundary to the top one in the static state. (With the fixed boundary temperatures we impose here, this temperature difference is unchanged by fluid motion.) The resulting dimensionless Boussinesq equations are \citep{Chandrasekhar1981}
\begin{align}
\nabla \cdot \mathbf u &= 0 \label{eq: inc} \\
\partial_t \mathbf u + \mathbf u \cdot \nabla \mathbf u &=
	-\nabla p + \Pran \nabla^2 \mathbf u + \Pran Ra\,T\,\hat{\mathbf z} \label{eq: u} \\
\partial_t T + \mathbf u \cdot \nabla T &= \nabla^2 T. \label{eq: T}
\end{align}
In our present 2D geometry, the velocity, $\bu=(u,w)$, and temperature, $T$, are defined on a dimensionless spatial domain bounded vertically by $-1/2\le z\le1/2$ and in the (periodic) horizontal direction by $0\le x < A$, where $A$ is the aspect ratio. We fix $A=2$ here, meaning the domain's width is twice its height. The two dimensionless control parameters are the Rayleigh and Prandtl numbers, defined respectively by
\begin{equation}
Ra = \frac{g\alpha d^3\Delta}{\kappa\nu} \qquad \Pran = \frac{\nu}{\kappa}, \label{eq: R}
\end{equation}
where $g$ is the gravitational acceleration in the $-\hat{\mathbf z}$ direction, and $\alpha$ is the linear coefficient of thermal expansion.

We have simulated the Boussinesq equations using the spectral element code {\tt nek5000} \citep{nek}, which integrates (\ref{eq: inc})-(\ref{eq: T}), and a fully spectral code that integrates the corresponding vorticity-stream function formulation. Our methods of computation and verification are described in Appendix \ref{app: comp}.

\subsection{Boundary conditions}
\label{sec: BC}

At the top and bottom boundaries, we impose free-slip velocity conditions,
\begin{equation}
w = \partial_z u = 0 \text{ at } z=\pm\tfrac{1}{2}, \label{eq: free-slip}
\end{equation}
so the boundaries exert no tangential stresses on the fluid. Boundaries in some astrophysical and plasma physical applications are indeed better modelled as free-slip than no-slip, though neither is quite realistic. Laboratory experiments are better modelled as no-slip, but we restrict ourselves to free-slip conditions because they admit strong zonal flows at computationally accessible $Ra$. When one boundary is changed to no-slip, the minimum $Ra$ needed for shearing convection can rise by several orders of magnitude, and it rises by even more when both boundaries are made no-slip \citep{VanderPoel2014a}. (With $Ra=10^9$, for instance, we have seen zonal flow between no-slip boundaries only when $A\lesssim1.4$. If shearing convection can occur in wider domains, it requires even larger $Ra$.)

The dimensionless boundary temperatures are fixed here as
\begin{equation}
T = \mp\tfrac{1}{2} \text{ at } z=\pm\tfrac{1}{2}. \label{eq: fixed-T}
\end{equation}
These temperatures could be translated arbitrarily since only their difference, which has been nondimensionalized to unity, affects the dynamics. Results of simulations with heat \emph{fluxes} fixed at the boundaries, carried out by us and by \citet{Zhu2012}, are similar to the fixed-temperature results described here, as is the case with non-shearing convection at large $Ra$ \citep{Johnston2009}.

\subsection{Integral quantities}
\label{sec: int quant}

To judge how zonal flow affects heat transport, we examine several integral quantities that are averaged over either the horizontal direction or the entire domain, and in some cases over time as well. We denote averages over the horizontal direction by an overbar and averages over the entire domain by angle brackets, as in
\begin{align}
\overline f(z,t) &:= \frac{1}{A}\int_0^{A} f(x,z,t)\,dx \label{eq: area} \\
\langle f \rangle(t) &:= \int_{-1/2}^{1/2} {\overline f}(z,t)\,dz. \label{eq: vol}
\end{align}
A superscript of $t$ indicates that the infinite-time average is also taken, as in $\overline{f}^t(z)$ and $\langle f\rangle^t$. In this notation, the time-averaged zonal flow profile is $\overline{u}^t(z)$. We define shearing convection as a flow in which $\overline{u}^t(z)$ is not identically zero. Non-shearing convection may have nonzero \emph{instantaneous} profiles, $\overline{u}(z,t)$, but these will vanish in long-time averages.

To quantify the strength of the zonal flow, we decompose the kinetic energy, $E$, into a horizontal part, $E_x$, and vertical part, $E_z$, where
\begin{align}
E_x&=\tfrac{1}{2}\langle u^2\rangle^t & 
E_z&=\tfrac{1}{2}\langle w^2\rangle^t &
E &= E_x+E_z.
\label{eq: E}
\end{align}
At large $Ra$, zonal flow accounts for almost all of the horizontal motion, so the zonal energy is thus roughly equal to the horizontal energy, $E_x$. The remaining horizontal energy, residing in nonzero horizontal wavenumbers, is comparable to the vertical energy, and they together comprise the non-zonal energy, which is roughly twice $E_z$.

Bulk heat transport in RB convection is typically quantified by the dimensionless Nusselt number, $N$, which is defined as the total heat flux due to both convection and conduction, normalized by that due to conduction alone. With the fixed-temperature boundary conditions we impose, the (dimensionless) mean conductive flux is fixed at unity, and the mean convective flux is $\langle wT\rangle^t$, so the spatiotemporally averaged Nusselt number is
\begin{equation}
N = 1 + \langle wT\rangle^t. \label{eq: N}
\end{equation}
The instantaneous Nusselt number is analogously defined as $N(t)=1+\langle wT\rangle(t)$.

Several scaling arguments have been put forth to explain and predict the parameter-dependence of $N(Ra,\Pran)$ in RB convection. The arguments of \citet{Grossmann2000} were perhaps the first to systematically predict a number of different scalings in different parameter regimes. In their most recent form, the arguments predict that $N$ will vary roughly as a power of $Ra$, with powers between $1/5$ and $1/2$ in various regimes \citep{Stevens2013}. As far as we know, all growth rates reported for past physical and numerical experiments fall within this range \citep{Grossmann2000, Ahlers2009}. In the present work, however, we find that zonal flow can cause $N$ to grow more slowly than $Ra^{1/5}$ and even, in some parameter regimes, to decrease as $Ra$ is raised.

\section{Sustained shearing convection at a large Prandtl number}
\label{sec: sustained}

\begin{figure}
\centering
(a)\fbox{\includegraphics[width=170pt]{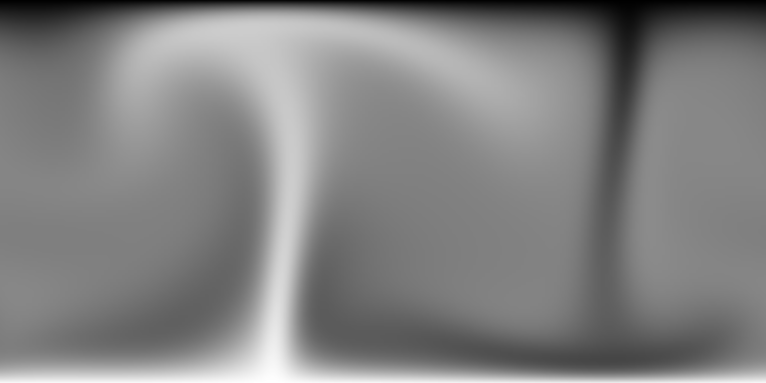}} \hspace{0pt}
(b)\fbox{\includegraphics[width=170pt]{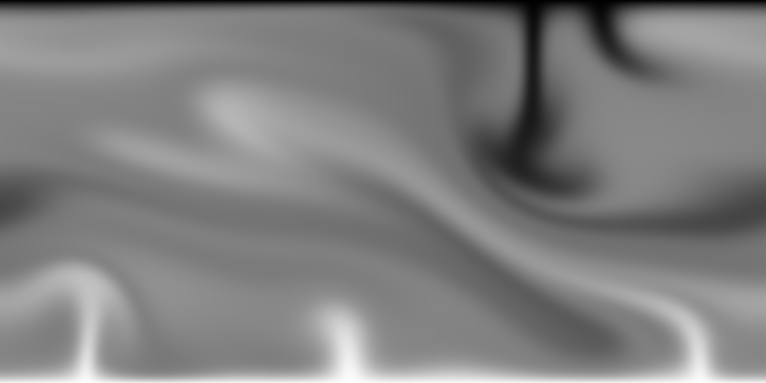}} \\ \vspace{8pt}
(c)\fbox{\includegraphics[width=170pt]{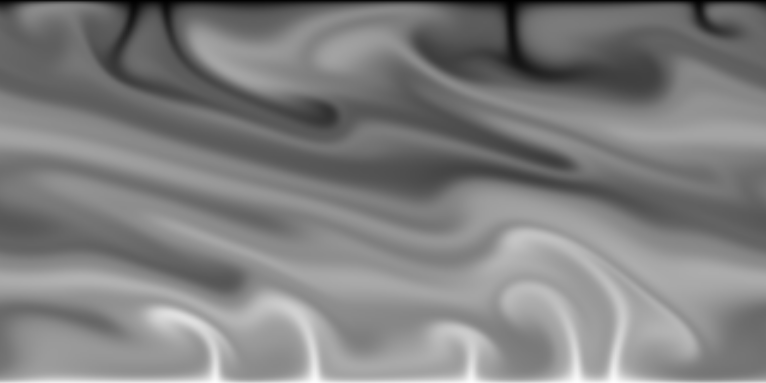}} \hspace{0pt}
(d)\fbox{\includegraphics[width=170pt]{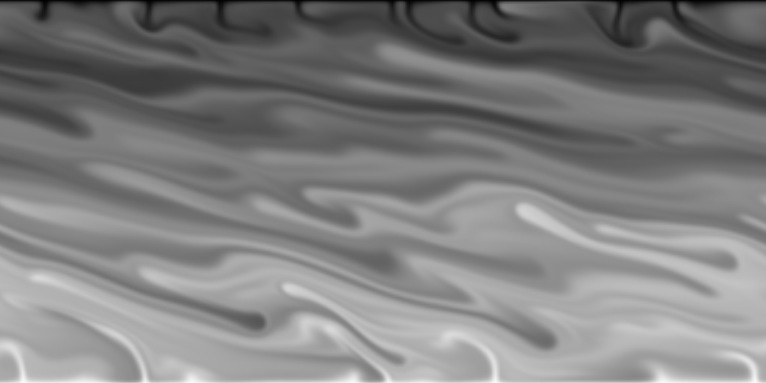}}
\caption{(Colour \href{http://www-personal.umich.edu/~goluskin/convection-driven-shear-videos}{videos} online) Instantaneous temperature fields with $\Pran=10$ and $Ra=2\e5$, $2\e6$, $2\e7$, and $2\e8$ (in a-d, respectively). The lowest-$Ra$ case ($a$) is non-shearing. The other three cases are examples of sustained shearing convection in which the zonal flow travels leftward along the top boundary and rightward along the bottom one. The hottest fluid (white) is one dimensionless degree warmer than the coldest fluid (black).}
\label{fig: video stills}
\end{figure}

Shearing convection is shown in the temperature fields of figures \hbox{\ref{fig: video stills}(b-d)} for $Ra=2\e6$, $2\e7$, and $2\e8$, and non-shearing convection is shown for comparison in figure \hbox{\ref{fig: video stills}(a)}. In the non-shearing state, counter-rotating rolls produce rising hot plumes and falling cold ones; these plumes fluctuate but have no mean drift to the left or right. In the shearing states of figures \hbox{\ref{fig: video stills}(b-d)}, and especially in the accompanying online \href{http://www-personal.umich.edu/~goluskin/convection-driven-shear-videos}{videos}, zonal flows are evidenced by the leftward drift of cold plumes along the top and the rightward drift of hot plumes along the bottom. Top and bottom boundary layers differ instantaneously but not in long-time averages. Although the flows of figures \hbox{\ref{fig: video stills}(b-d)} are all sheared in the same way, their mirror images under the reflection $x\mapsto-x$ can arise just as easily; the reflectional symmetry of the governing equations can be broken in either sense. Once the zonal flow has fully developed with one sign or the other, however, we have never seen it reverse at large $Ra$, nor were \citet{Krishnamurti1981} able to induce such reversals in experiments. 

The temperature fields of shearing convection in figures \hbox{\ref{fig: video stills}(b-d)} suggest how zonal flows can persist despite not being directly driven by buoyancy forces. As the fluid in a hot plume rises, it is diverted leftward, first by the sheared mean flow and then by the top boundary, feeding the zonal flow in the upper part of the domain. Likewise, the fluid in a cold plume is diverted rightward as it falls. Said another way, energy created by the action of buoyancy forces cascades upscale, via the velocity nonlinearity, and into the zonal flow. The transfer of energy into the zonal flow is stronger at larger~$Ra$, smaller~$\Pran$, and smaller~$A$ \citep{Howard1986, Rucklidge1996, GoluskinThesis}. These parameter trends are reflected in the linear instability of steady rolls to perturbations of the zonal flow \citep{Thompson1970, Busse1983, Howard1986}, and they hold also when the zonal flow is dominant, as in the convection studied here.

\subsection{Parameter regimes and bistability}

\begin{figure}
\centering
\includegraphics[width=260pt]{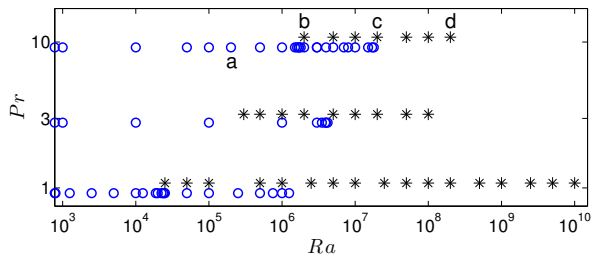}
\caption{Rayleigh number ranges over which we have found shearing ($*$) and non-shearing ({\color{blue} {\large$\circ$}}) convection to persist with $A=2$ and $Pr=1$, 3, or 10. Upper bounds on the non-shearing regimes ($Ra=1.2\e6$, $3.5\e6$, and $1.8\e7$) and the lower bounds on the shearing regimes ($Ra=2.5\e4$, $3\e5$, and $2\e6$) are only approximate. The shearing flows are \emph{bursting} when $\Pran=1$ and \emph{sustained} when $\Pran=3$ or 10 (see text).}
\label{fig: regimes}
\end{figure}

Figure \ref{fig: regimes} summarizes the parameter values at which we have simulated shearing states ($*$) and non-shearing states ({\color{blue} {\large$\circ$}}) that persist in a domain of aspect ratio 2. The points labeled $a$-$d$ correspond to the flows shown in figure \ref{fig: video stills}. At each of the three $\Pran$ simulated, there is a bistable regime in which both shearing and non-shearing states can persist. Non-shearing states in the bistable range of $Ra$ can be found by beginning with $Ra$ below the range and slowly raising it, while shearing states can be found by beginning with $Ra$ above the range and slowly lowering it. It is hard to pinpoint the $Ra$ values at which the various solution branches lose stability because transients can be very long, but the data suggest two conclusions---first, that shearing convection is inevitable as $Ra\to\infty$, and, second, that raising $Pr$ also raises the $Ra$ at which the transition to shearing convection occurs.

In addition to distinguishing shearing convection from its non-shearing counterpart, we can distinguish between the \emph{sustained shearing convection} that occurs at larger $\Pran$ and the \emph{bursting shearing convection} that occurs at smaller $\Pran$. We refer to the sustained flows as such because convective transport is significant at all times, meaning that the instantaneous Nusselt number, $N(t)$, never drops close to unity. We refer to the bursting flows as such because convective transport occurs almost exclusively during discrete bursts, in between which $N(t)$ is very close to unity. The distinction between the sustained and bursting types of shearing convection is qualitative, but there is ambiguity only over a narrow transitional range of Prandtl numbers, as described in \S\ref{sec: transition}. The present section focuses on sustained shearing convection with $Pr=10$, of which figures \hbox{\ref{fig: video stills}(b-d)} are examples. The shearing convection we have simulated is sustained also when $Pr=3$, whereas it is bursting when $Pr=1$. Bursting shearing convection is the focus of \S\ref{sec: bursting}.

The bistable regime shrinks and disappears as the imposed horizontal period, $A$, is decreased. This is consistent with the bifurcation structure of steady states, wherein shearing states arise via supercritical bifurcations when $A$ and $\Pran$ are sufficiently small \citep{Hermiz1995, GoluskinThesis, Fitzgerald2014}. The narrower periods imposed in most prior studies of the full PDEs \citep{Rucklidge1996, Garcia2003, Garcia2003a, Bian2003a} seem to be too narrow for there to be bistability between shearing and non-shearing convection.

\subsection{Comparison of integral quantities in shearing and non-shearing convection}
\label{sec: comparison}

\begin{figure}
\centering
(a)\includegraphics[width=178pt]{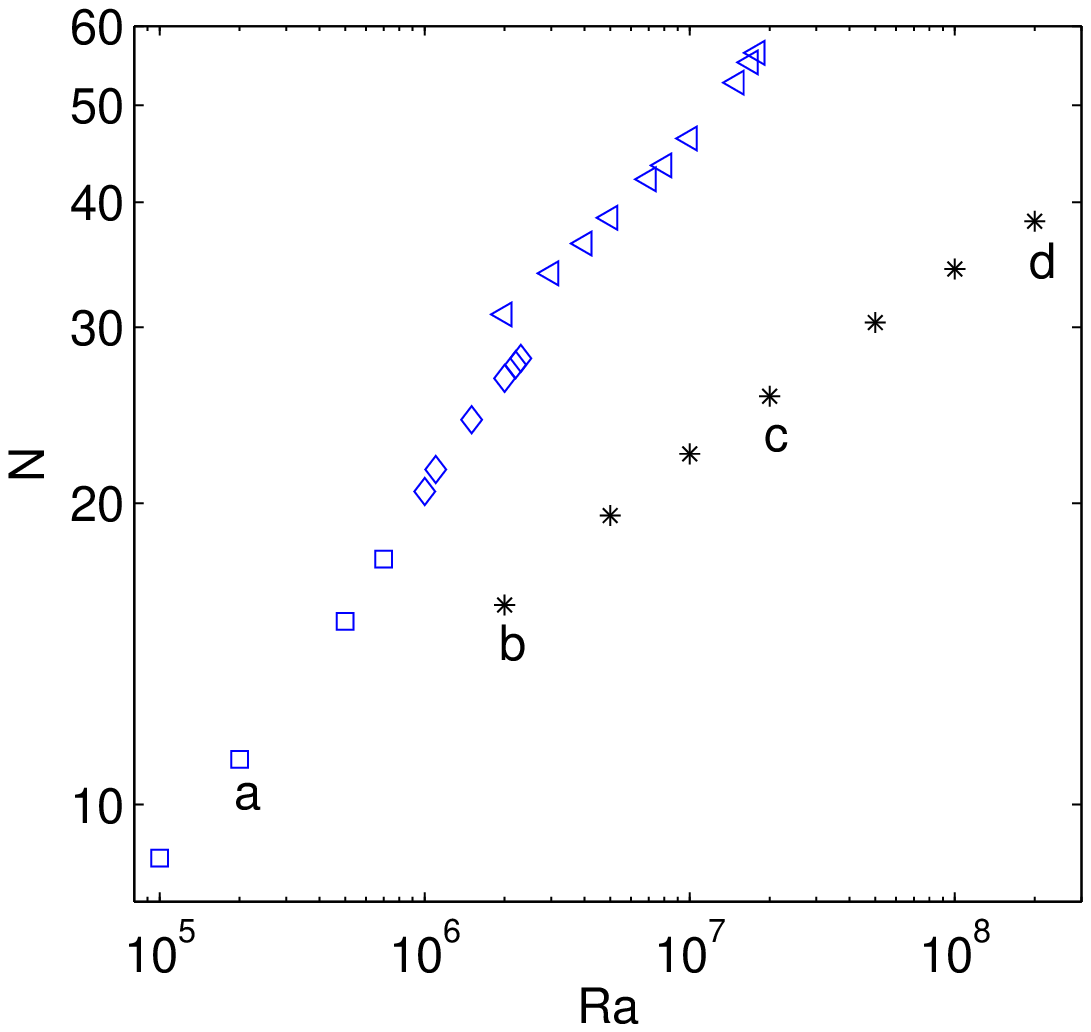} \\ 
(b)\includegraphics[width=178pt]{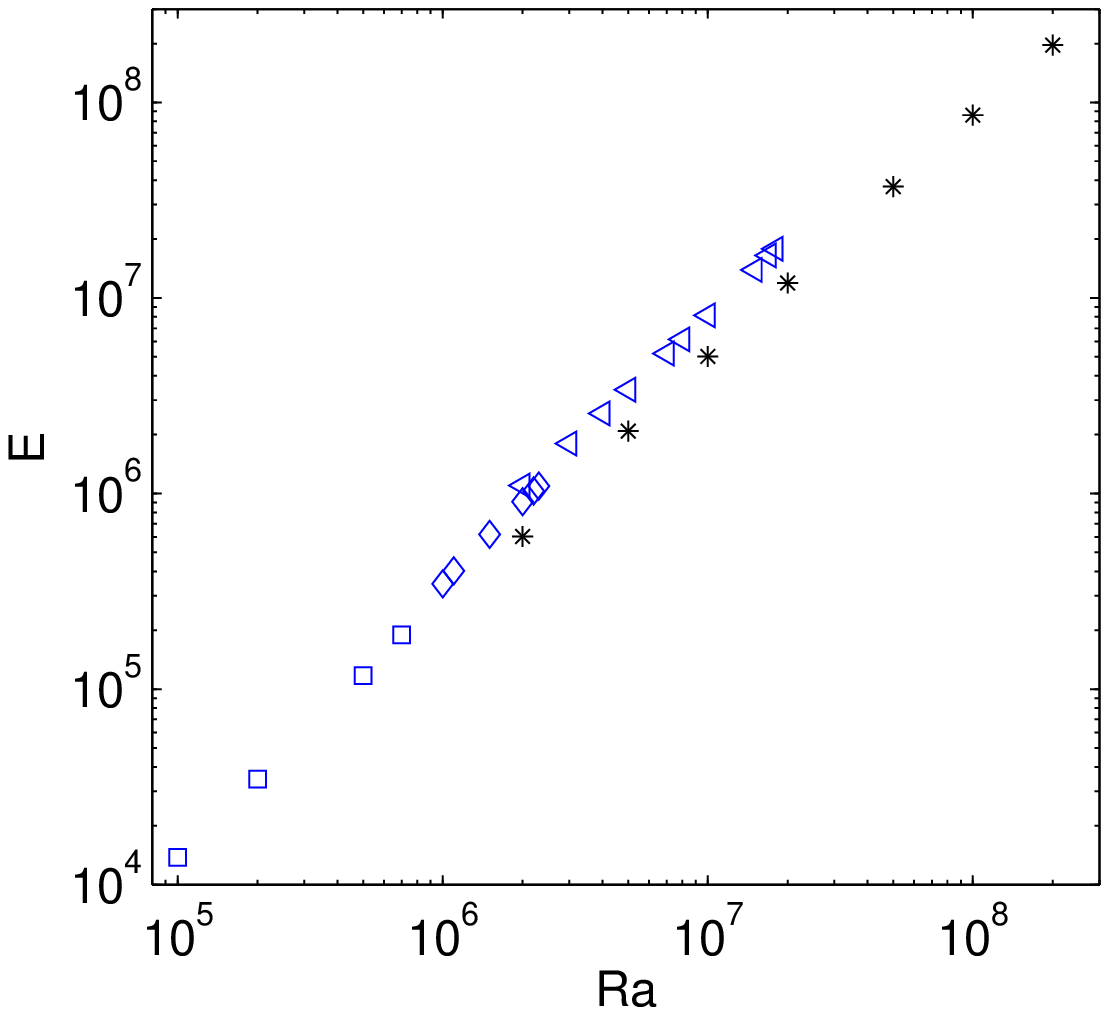} 
(c)\includegraphics[width=178pt]{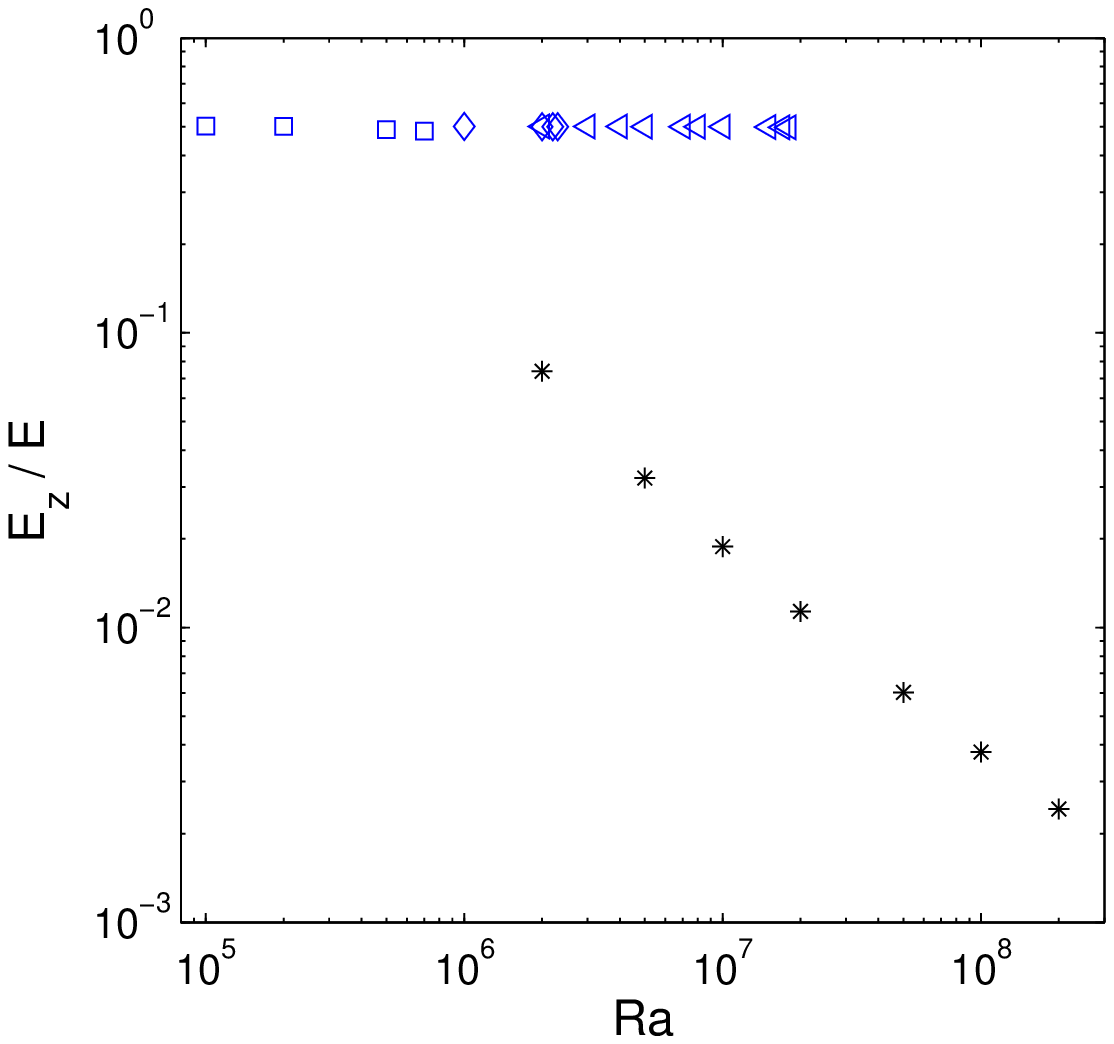}
\caption{Time-averaged Nusselt numbers ($N$), total kinetic energies ($E$), and the fractions of kinetic energies due to vertical velocities ($E_z/E$) in sustained shearing convection ($*$) and non-shearing convection at various Rayleigh numbers for $(A,\Pran)=(2,10)$. The non-shearing solutions seem to be variously chaotic (\textcolor{blue}{\tiny{$\square$}}), periodic (\textcolor{blue}{\small{$\diamond$}}), and quasiperiodic (\textcolor{blue}{\small{$\triangleleft$}}), while the shearing solutions all seem to be chaotic. Temperature fields from the cases labelled $a$-$d$ are shown in figures \hbox{\ref{fig: video stills}(a-d)}, respectively.}
\label{fig: Pr=10 int}
\end{figure}

For both shearing and non-shearing solutions, figure \ref{fig: Pr=10 int} shows the Rayleigh-number dependence of time-averaged Nusselt numbers ($N$), total kinetic energies ($E$), and the fractions of these kinetic energies contributed by vertical velocities ($E_z/E$). We believe the non-shearing states are variously chaotic, periodic, and quasiperiodic, as indicated in the figure, while the shearing states are all chaotic at this Prandtl number. (Throughout this work, we refer to solutions as quasiperiodic when multiple incommensurable temporal frequencies are present and as chaotic when they appear to have broken all spatial and temporal symmetries of the governing PDEs. We have not demonstrated that any mathematical definition of chaos is met.) The solutions represented in the temperature fields of figures \hbox{\ref{fig: video stills}(a-d)} are labelled $a$-$d$, respectively, in figure \hbox{\ref{fig: Pr=10 int}(a)}.

Some integral quantities, but not all, are significantly affected by zonal flow. Nusselt numbers are smaller and increase more slowly with $Ra$ in shearing convection than in non-shearing convection at the same parameters---the Nusselt numbers of figure \hbox{\ref{fig: Pr=10 int}(a)'s} shearing states ($*$) grow approximately like $Ra^{0.19}$, while those of its chaotic non-shearing states (\textcolor{blue}{\scriptsize{$\square$}}) grow like $Ra^{0.35}$. (Heat transport in shearing convection is discussed further in \S5.) Kinetic energies, $E$, grow at similar rates in shearing and non-shearing convection. This is shown for $\Pran=10$ in figure \hbox{\ref{fig: Pr=10 int}(b)} and remains true in our simulations at other Prandtl numbers. However, the fractions of the kinetic energy in the vertical motion, $E_z/E$, differ greatly between the two types of flows. In non-shearing convection, horizontal and vertical motions are comparable at all $Ra$, so the fractions remain near 1/2. In shearing convection, $E_z/E$ is already smaller than 1/10 when $Ra=2\e6$, and it decreases further as $Ra$ is raised, even though $E_z$ itself increases. If this trend continues at very large $Ra$, after the onset of shear turbulence, then $E_z/E$ will approach zero as $Ra\to\infty$. In other words, the fraction of kinetic energy in the zonal flow will approach unity.

Energy fluxes shed light on how zonal flow can so greatly change the ways that $N$ and $E_z/E$ depend on $Ra$ while hardly changing the dependence of $E$ on $Ra$. Averaging $\bu\cdot(\ref{eq: u})$ over space and time yields the time-averaged velocity power integral \citep{Malkus1954a, Howard1963},
\begin{equation}
Ra(N-1) = \langle |\nabla\bu|^2\rangle^t. \label{eq: u power int}
\end{equation}
The left-hand side of expression (\ref{eq: u power int}) is the mean rate at which the flow gains kinetic energy from buoyancy forces (in suitable units), while the right-hand side is the mean rate at which it loses energy to viscous dissipation. Being equal in the infinite-time limit, either flux can be thought of as the mean energy throughput. Although kinetic energies, $E$, grow with $Ra$ at very similar rates in shearing and non-shearing convection, the energy throughput grows more slowly in the shearing case. This is possible because the kinetic energy of shearing convection resides, on average, at larger scales and so dissipates more slowly, meaning the right-hand side of (\ref{eq: u power int}) is smaller. Correspondingly, less work by buoyancy is needed to maintain the kinetic energy, hence the Nusselt number on the left-hand side of (\ref{eq: u power int}) is smaller.

\subsection{Spatial structure of sustained shearing convection}
\label{sec: sustained various R}

The vertical structure of sustained shearing convection is markedly different from that of non-shearing convection. In both cases, the thermal boundary layers become thinner as $Ra$ is raised, meaning more heat is being conducted upward across them, and the plumes shrink proportionally while becoming more numerous. But, it seems, only in non-shearing convection is the shrinking of the plumes accompanied by their increasing tendency to cluster into composite plumes that can penetrate the whole interior \citep{Parodi2004, VonHardenberg2008}. It is evident from the temperature fields of figures \hbox{\ref{fig: video stills}(b-d)} that there is no such clustering in sustained shearing convection. Consequently, as $Ra$ is raised, the plumes penetrate less and less far before losing coherence. This is because the mean shear is growing stronger and thus more effective at dispersing plumes horizontally---a phenomenon known as Taylor shear dispersion \citep{Taylor1953}. The dispersion of plumes in shearing convection is more involved than in \citeauthor{Taylor1953}'s study because the shear, rather than being imposed \emph{a priori}, is dynamically driven via the very plumes it helps disperse. When the plumes exceed their time-averaged strength, the zonal flow strengthens and further disperses them; when the plumes fall below this strength, the zonal flow weakens and lets them grow.

\begin{figure}
\centering
(a)\includegraphics[height=150pt]{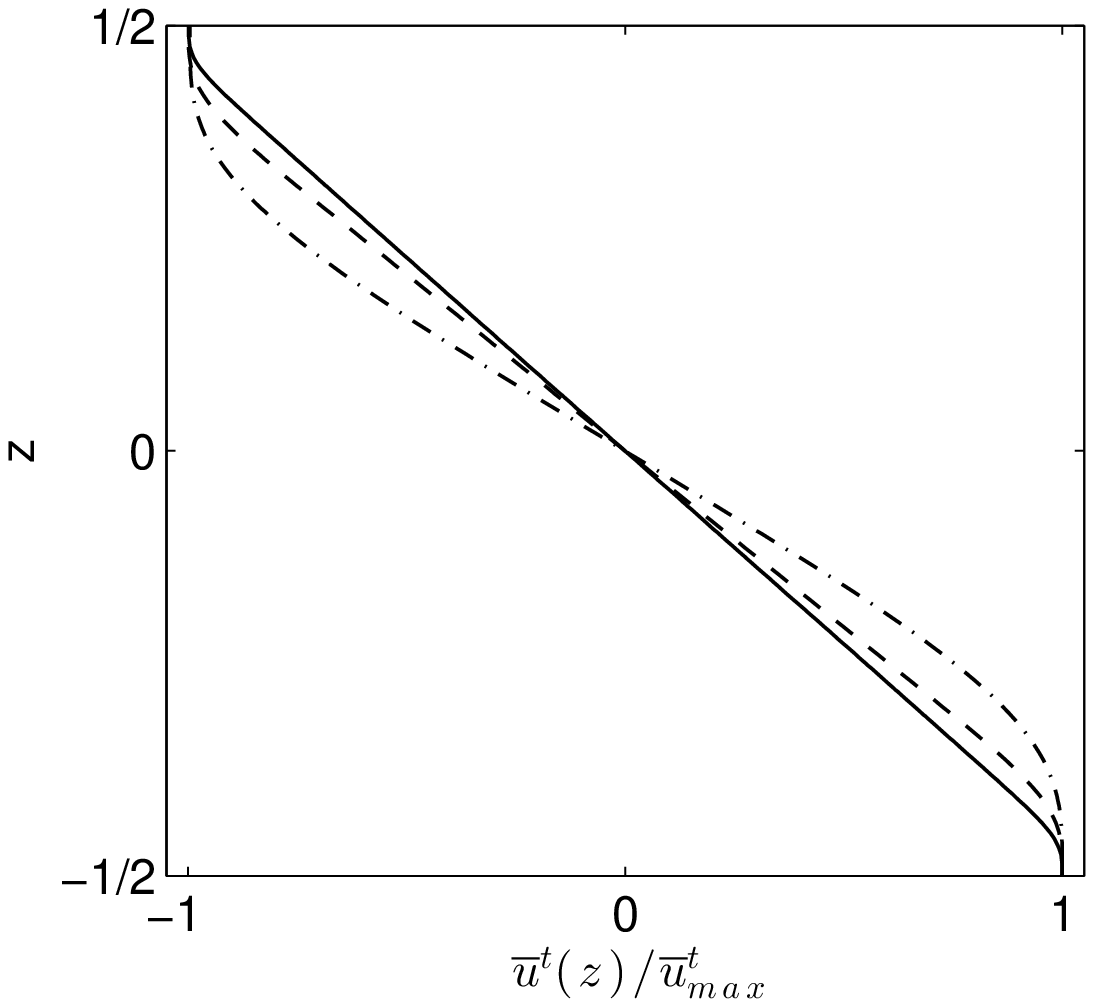} \hspace{8pt}
(b)\includegraphics[height=150pt]{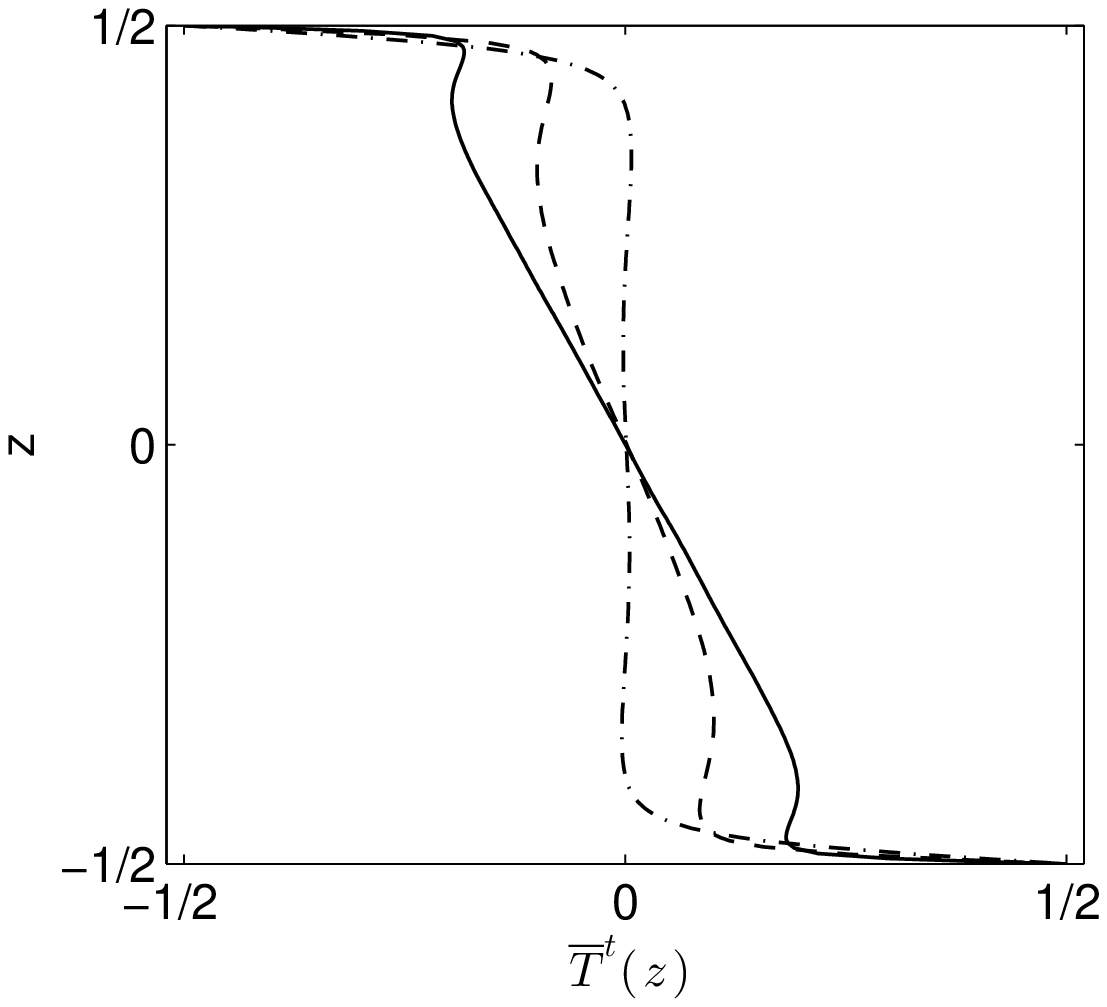}
\caption{Mean vertical profiles of zonal flow (left) and temperature (right) for sustained shearing convection with $(A,\Pran)=(2,10)$ at Rayleigh numbers of $2\cdot10^6$ (\protect\dashdotrule), $2\cdot10^7$ (\protect\dashedrule), and $2\cdot10^8$ (\solidrule). Each zonal flow profile is normalized by its maximum value, $\overline{u}^t_{max}$. Reversing $\overline{u}^t(z)$ yields the profiles that arise when symmetry breaks in the opposite way. Figure \ref{fig: Pr=1 profiles} shows analogous profiles for \emph{bursting} shearing convection.}
\label{fig: Pr=10 profiles}
\end{figure}

For the three examples of shearing convection visualized in figures \hbox{\ref{fig: video stills}(b-d)}, mean vertical profiles of the normalized horizontal velocity, $\overline{u}^t(z)/\overline{u}^t_{max}$, and the temperature, $\overline{T}^t(z)$, are shown in figure \ref{fig: Pr=10 profiles}. All three velocity profiles are roughly linear in the interior, so the shearing of smaller-scale structures by the zonal flow is roughly constant in $z$. This interior region is surrounded by boundary layers in which the shear stresses vanish, as required by the free-slip conditions, and these  boundary layers shrink as $Ra$ is raised.

Mean temperature profiles in shearing convection are more structured than their non-shearing counterparts. Of the three shearing states represented in figure \hbox{\ref{fig: Pr=10 profiles}(b)}, only at the lowest $Ra$, where the zonal flow is weakest, does the $\overline{T}^t(z)$ profile resemble those of non-shearing convection. That is, there are two thermal boundary layers where heat is transported almost solely by conduction and a nearly isothermal interior where heat is transported almost solely by convection. On the other hand, for the states with $Ra=2\e7$ and $2\e8$, in which the zonal flow is more dominant, we can identify five distinct regions that are delimited by changes in the sign of $\frac{d\overline{T}^t}{dz}(z)$: two conductive thermal boundary layers, two mixing layers, and the interior.

To interpret the alternating signs of $\frac{d\overline{T}^t}{dz}(z)$ in the profiles of figure \hbox{\ref{fig: Pr=10 profiles}(b)} with $Ra=2\e7$ and $2\e8$, we recall that the time-averaged heat flux is the same across every horizontal plane; only the relative contributions of conduction, $-\frac{d\overline{T}^t}{dz}(z)$, and convection, $\overline{wT}^t(z)$, can change with height, so
\begin{equation}
N=-\tfrac{d\overline{T}^t}{dz}(z)+\overline{wT}^t(z) \label{eq: N(z)}
\end{equation}
for all $z$. Just as in non-shearing convection, heat transport in the thermal boundary layers is almost totally conductive. In the mixing layers, defined as the regions where $\frac{d\overline{T}^t}{dz}(z)>0$, conduction actually carries heat downward but is overmatched by strong upward convection. The vertical extent of these mixing layers roughly corresponds to regions where the plumes are pronounced, as can be seen by comparing figures \hbox{\ref{fig: video stills}(c-d)} and \hbox{\ref{fig: Pr=10 profiles}(b)}. In the interior, convection and conduction both contribute to upward heat transport. Some sheared plumes are still discernible there, but they transport less heat than in the mixing layers.

Convection has been studied in which large-scale shear is driven by counter-moving boundaries, rather than by the convection itself. Like the spontaneously arising shear we describe, this imposed shear depresses heat transport at large $Ra$ \citep{Lipps1971, Doamaradzki1988, Zaleski1991}. Temperature profiles reported for systems with counter-moving boundaries resemble their counterparts in non-shearing convection, rather than the more structured profiles we have found with convectively driven shear, but this would likely change with stronger imposed shear and larger $Ra$. For Poiseuille-Rayleigh-B\'enard convection, where shear is induced by a horizontal pressure gradient, \citet{Scagliarini2014} report temperature profiles that lack inversions but resemble our own in the interior region.

\subsection{Transition between the sustained and bursting regimes of shearing convection}
\label{sec: transition}

\begin{figure}
\centering
\begin{tabular}{ccc}
\qquad $\Pran=10$:& 
\qquad $\Pran=5$:& 
\qquad $\Pran=3$:\vspace{4pt}\\
\includegraphics[height=100pt]{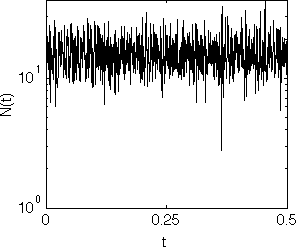} &
\includegraphics[height=100pt]{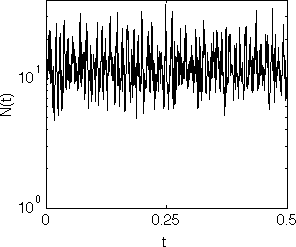} &
\includegraphics[height=100pt]{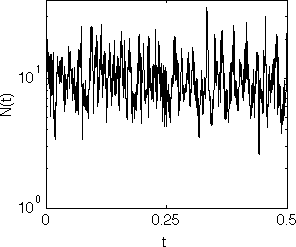} \vspace{6pt} \\
\qquad $\Pran=2$: &
\qquad $\Pran=1.5$: &
\qquad $\Pran=1$: \vspace{4pt} \\
\includegraphics[height=100pt]{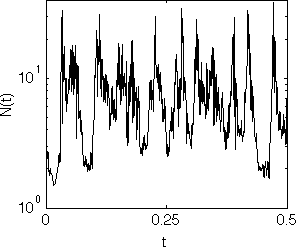} & 
\includegraphics[height=100pt]{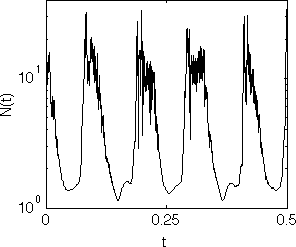} &
\includegraphics[height=100pt]{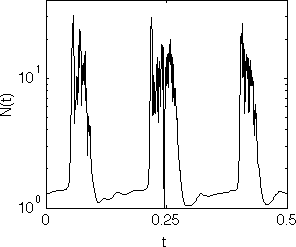}
\end{tabular}
\caption{Time series of instantaneous (volume-averaged) Nusselt numbers in shearing convection for six Prandtl numbers between 1 and 10 with $(A,Ra)=(2,2\cdot10^6)$. The transition between the sustained and bursting versions of shearing convection is illustrated.}
\label{fig: Ra=2e6}
\end{figure}

The qualitative distinction between the sustained shearing convection described thus far and the bursting shearing convection described in \S\ref{sec: bursting} is evident in \href{http://www-personal.umich.edu/~goluskin/convection-driven-shear-videos}{videos} of evolving temperature fields (available online). It is also evident in time series of various quantities, such as the instantaneous Nusselt number, $N(t)$. Figure \ref{fig: Ra=2e6} shows the onset of bursting in $N(t)$ as the Prandtl number is lowered from 10 to 1 with $Ra=2\e 6$. Raising the Prandtl number back through the same values produces similar time series; the transition to bursting seems to be smooth and without hysteresis. At the largest three Prandtl numbers represented in figure \ref{fig: Ra=2e6}, $N(t)$ remains well above unity at all times, meaning the vertical convection that drives the flow is always significant. This is the sense in which the convection is \emph{sustained}. Similarly, time series such as $E_x(t)$ and $E_z(t)$ in sustained shearing convection deviate only modestly from their time-averaged values (cf.\ Appendix~\ref{app: time series}).

The fluctuations of $N(t)$ in figure \ref{fig: Ra=2e6} intensify as $\Pran$ is lowered until, when $\Pran$ is near 1.5, they become so pronounced that $N(t)$ begins to ``bottom out'' near unity. This creates a quiescent phase with little convective transport, punctuated by convective bursts---\emph{bursting shearing convection}. The transition occurs over a fairly narrow range of $\Pran$ but not at a well-defined value. Quiescent phases are hinted at (in figure \ref{fig: Ra=2e6}) when $\Pran=2$, are short when $\Pran=1.5$, and are quite pronounced when $\Pran=1$. The transition between bursting and sustained shearing convection thus occurs over the approximate range $1.5\lesssim\Pran\lesssim2$ when $Ra=2\e6$. This range is similar at larger $Ra$, at least up to~$10^8$.

\section{Bursting shearing convection at a moderate Prandtl number}
\label{sec: bursting}

In this section, we examine \emph{bursting shearing convection}, which we have simulated with $(A,\Pran)=(2,1)$ and Rayleigh numbers in the range $2.5\e4\le Ra\le10^{10}$. Unlike the \emph{sustained} shearing convection described in \S\ref{sec: sustained}, bursting flows at a Prandtl number of unity have been simulated previously \citep{Garcia2003, Garcia2003a, Bian2003a, VanderPoel2014a}. Moreover, exactly periodic bursts have been seen in reduced models that account for zonal flow \citep{Leboeuf1993, Malkov2001, Bian2003}, and less pronounced bursts can occur even when the large-scale shear is driven directly by boundary motion \citep{Zaleski1991}. We add to past findings here by simulating a wider range of Rayleigh numbers in a larger domain.

\subsection{The physics of a burst}

\begin{figure}
\begin{tabular}{rrr}
$Ra=2\cdot10^6$: \hspace{25pt} &
$Ra=2\cdot10^7$: \hspace{25pt} &
$Ra=2\cdot10^8$:  \hspace{25pt}\vspace{5pt} \\
\includegraphics[width=122.5pt]{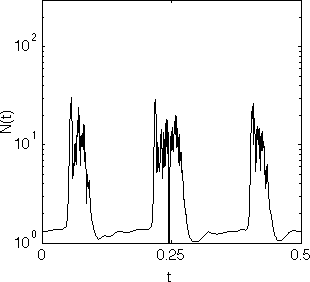} &
\includegraphics[width=122.5pt]{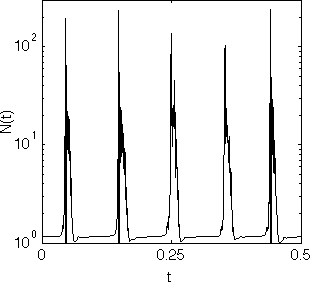} &
\includegraphics[width=122.5pt]{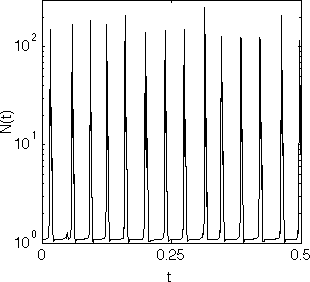} \vspace{2pt}\\
\includegraphics[width=116.5pt]{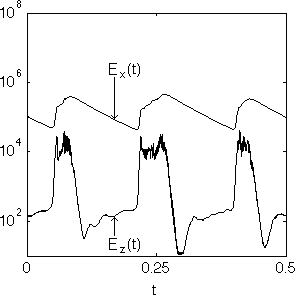} &
\includegraphics[width=116.5pt]{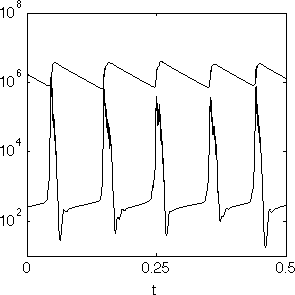} &
\includegraphics[width=116.5pt]{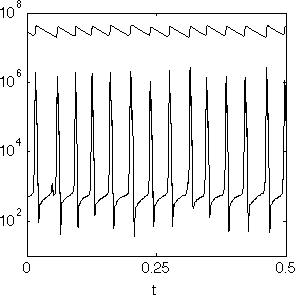}
\end{tabular}
\caption{Time series of the Nusselt numbers (top row) and the kinetic energies (bottom row) of the horizontal velocities ($E_x$, upper curves) and vertical velocities ($E_z$, lower curves) in bursting shearing convection with $(A,\Pran)=(2,1)$. The three Rayleigh numbers represented are $2\cdot10^6$ (left column), $2\cdot10^7$ (centrecolumn), and $2\cdot10^8$ (right column). The integral quantities are as defined in expressions (\ref{eq: E}) and (\ref{eq: N}) but with time averages not taken.}
\label{fig: Pr=1}
\end{figure}

\begin{figure}
\centering
\includegraphics[width=310pt]{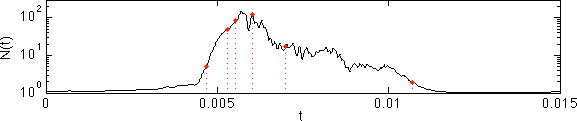} \vspace{2pt} \\
\includegraphics[width=310pt]{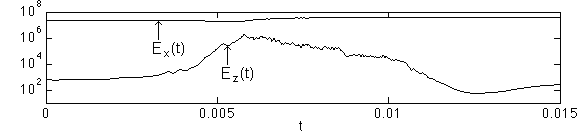} \vspace{12pt} \\
(a)\fbox{\includegraphics[width=150pt]{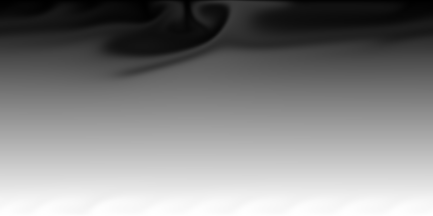}}
\hspace{0pt}
(b)\fbox{\includegraphics[width=150pt]{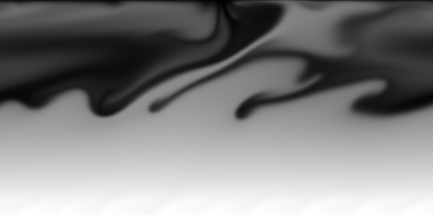}} \\
\vspace{8pt}
(c)\fbox{\includegraphics[width=150pt]{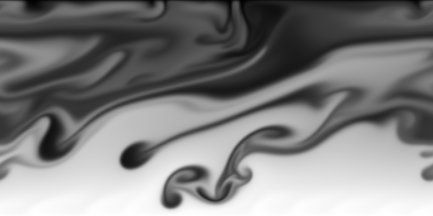}}
\hspace{0pt}
(d)\fbox{\includegraphics[width=150pt]{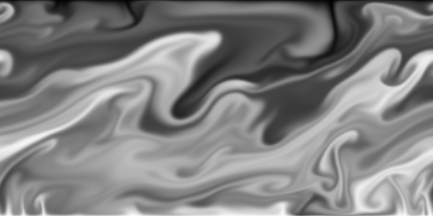}} \\
\vspace{8pt}
(e)\fbox{\includegraphics[width=150pt]{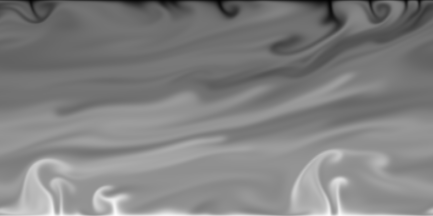}}
\hspace{0pt}
(f)\fbox{\includegraphics[width=150pt]{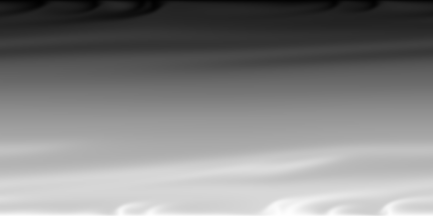}}
\caption{(Colour \href{http://www-personal.umich.edu/~goluskin/convection-driven-shear-videos}{video} online) Time series of $N(t)$, $E_x(t)$, and $E_z(t)$ during a single burst, along with six instantaneous temperature fields, with $(\Pran,Ra)=(1,2\e8)$. The time origin is arbitrary, and the full quiescent phases are not shown. The instants represented in fields $a$-$f$ are marked from left to right, respectively, by dashed lines on the time series of $N(t)$. The hottest fluid (white) is one dimensionless degree warmer than the coldest fluid (black).}
\label{fig: burst}
\end{figure}

Figure \ref{fig: Pr=1} shows time series of $N(t)$, $E_x(t)$, and $E_z(t)$ for bursting shearing convection with $Ra=2\cdot10^6$, $2\cdot10^7$, and $2\cdot10^8$. These are the same three Rayleigh numbers as in \S\ref{sec: sustained}'s examples of sustained shearing convection; only the Prandtl number has changed (from 10 to 1), as described in \S\ref{sec: transition}. In each time series of figure \ref{fig: Pr=1}, the quiescent phase is punctuated by global bursts that are much stronger than the flow's local fluctuations and have a much lower mean frequency. All three quantities in figure \ref{fig: Pr=1} rise quickly during a burst, after which $N(t)$ and $E_z(t)$ return quickly to their quiescent values, while $E_x(t)$ relaxes until the next burst. For the case with $Ra=2\e8$, the life cycle of a single burst is illustrated in figure \ref{fig: burst}, where $N(t)$ is shown alongside six representative temperature fields.

The physics behind the bursts of figure \ref{fig: Pr=1} has been discussed before \citep{Leboeuf1993, Garcia2006} and can be interpreted with the help of the three integral quantities shown. During the quiescent phase, there is little convective motion, so $N(t)\approx1$ and $E_z(t)\ll E_x(t)$. The small Nusselt number means that buoyancy is doing very little work on the fluid, so $E_x(t)$ relaxes towards zero. As convective structures are dispersed more and more, the vertical temperature profile comes very close to being linear, and the gravest zonal mode, $u\propto\sin\pi z$, dominates the flow because it decays the most slowly. That is,
\begin{align}
T &\approx -z \label{eq: relaxation T} \\
(u,w) &\approx (ce^{-\pi^2\Pran\,t}\sin \pi z,0), \label{eq: relaxation u} 
\end{align}
where $c$ is a constant that depends on the time origin and the parameters. At the large Rayleigh numbers under study, this conductive temperature profile would be highly unstable if the fluid were at rest. When the zonal flow has significant momentum, however, it suppresses the convective instability, so convection does not return until the zonal flow has decayed sufficiently. This same mechanism raises the critical Rayleigh number of linear instability when steady shear is imposed on the two-dimensional RB model \citep{Gallagher1965, Deardorff1965, Ingersoll1966}.

Bursts begin when the zonal flow becomes too weak to suppress the usual mechanism of convective instability, at which time $N(t)$ and $E_z(t)$ quickly grow, and the fluid overturns, as shown in figures \hbox{\ref{fig: burst}(b-d)} and accompanying online \href{http://www-personal.umich.edu/~goluskin/convection-driven-shear-videos}{videos}. Thermal plumes return, and the fluid is once again energized as buoyancy drives circulation. However, the newfound kinetic energy is quickly transferred into the zonal flow, which then shuts off the $N(t)$ and $E_z(t)$ bursts as a new quiescent phase begins.

To see the bursting from the perspective of dynamical systems, we can note that the relaxation defined by expressions (\ref{eq: relaxation T}) and (\ref{eq: relaxation u}) solves the governing PDEs exactly and lies in the stable manifold of the static state. During the quiescent phase, the flow is close to this exact solution. Each burst carries the system away from the relaxing solution and then back towards it, leaving the flow with more energy (that is, farther from the static equilibrium). Similar behaviour has been called \emph{on-off intermittency} \citep{Platt1993}, one of several types of fluid dynamical bursting surveyed by \citet{Knobloch1999a}.

Yet another way to view bursting shearing convection is by analogy with the oscillations of predator and prey populations in the Lotka-Volterra equations, a pair of first-order ODEs \citep{Lotka1925}. Several authors have spoken in these terms, and a few have even derived predator-prey-type ODEs from the Boussinesq equations under various closure assumptions \citep{Leboeuf1993, Malkov2001, Bian2003}.

\subsection{Dependence of bursts on the Rayleigh number}

The bursts shown in figure \ref{fig: Pr=1} change in several ways as $Ra$ is raised. They become more frequent, relative to the timescale of thermal diffusion by which we have nondimensionalized, and more extreme. In our simulations of \emph{sustained} shearing convection, on the other hand, $N(t)$ and $E_z(t)$ fluctuate less strongly as $Ra$ is raised, and the fluctuations in $E_x(t)$ are scarcely perceptible (cf.\ Appendix \ref{app: time series}).

The dependence of the time-averaged Nusselt number, $N$, on $Ra$ is not obvious from the time series of $N(t)$ in figure \ref{fig: Pr=1}. Raising $Ra$ makes the bursts of $N(t)$ stronger and more frequent but also shorter-lived, and these trends are in opposition. As reported in \S\ref{sec: nusselt}, the trends are in close competition, so $N$ changes only very gradually over several decades of $Ra$.

\subsection{Vertical and temporal structure}

\begin{figure}
\centering
(a)\includegraphics[width=170pt]{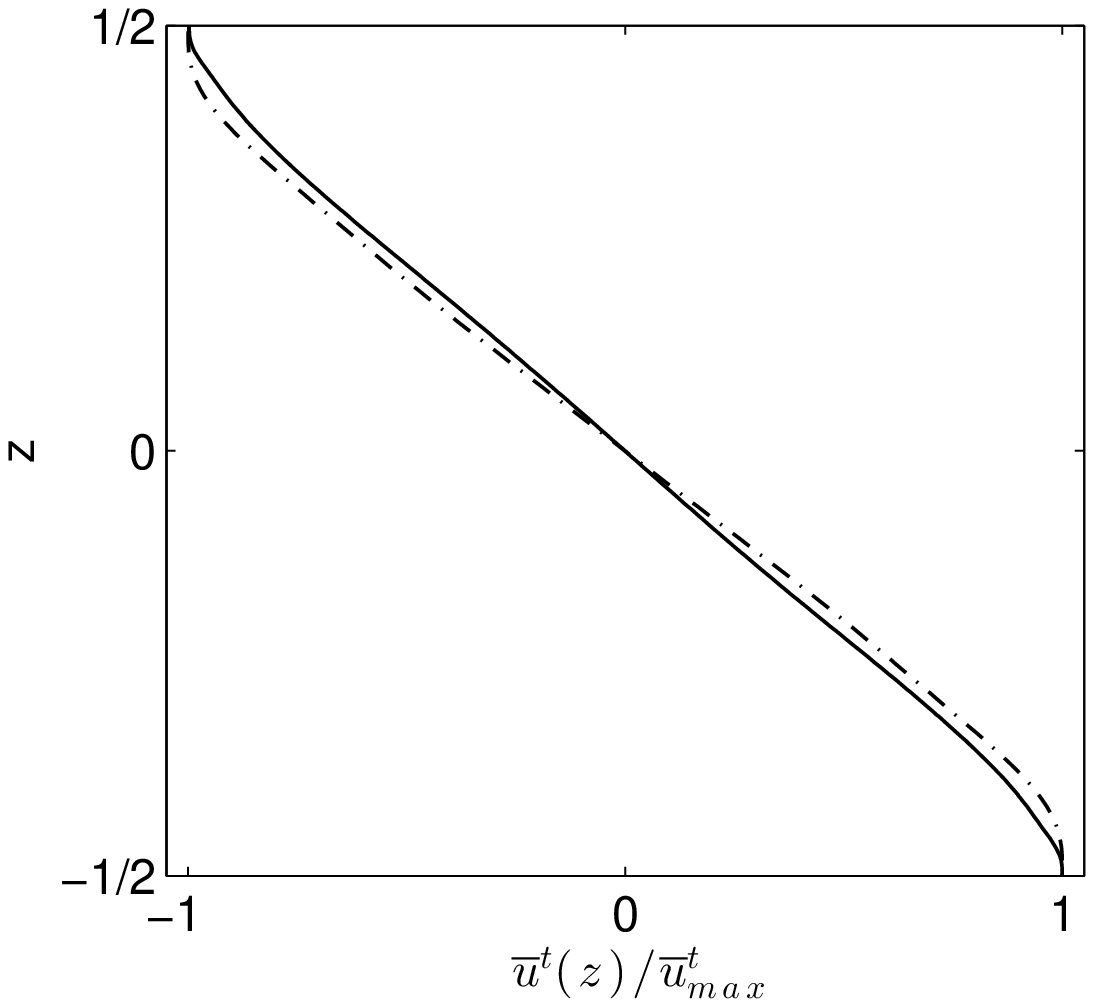} \hspace{8pt}
(b)\includegraphics[width=170pt]{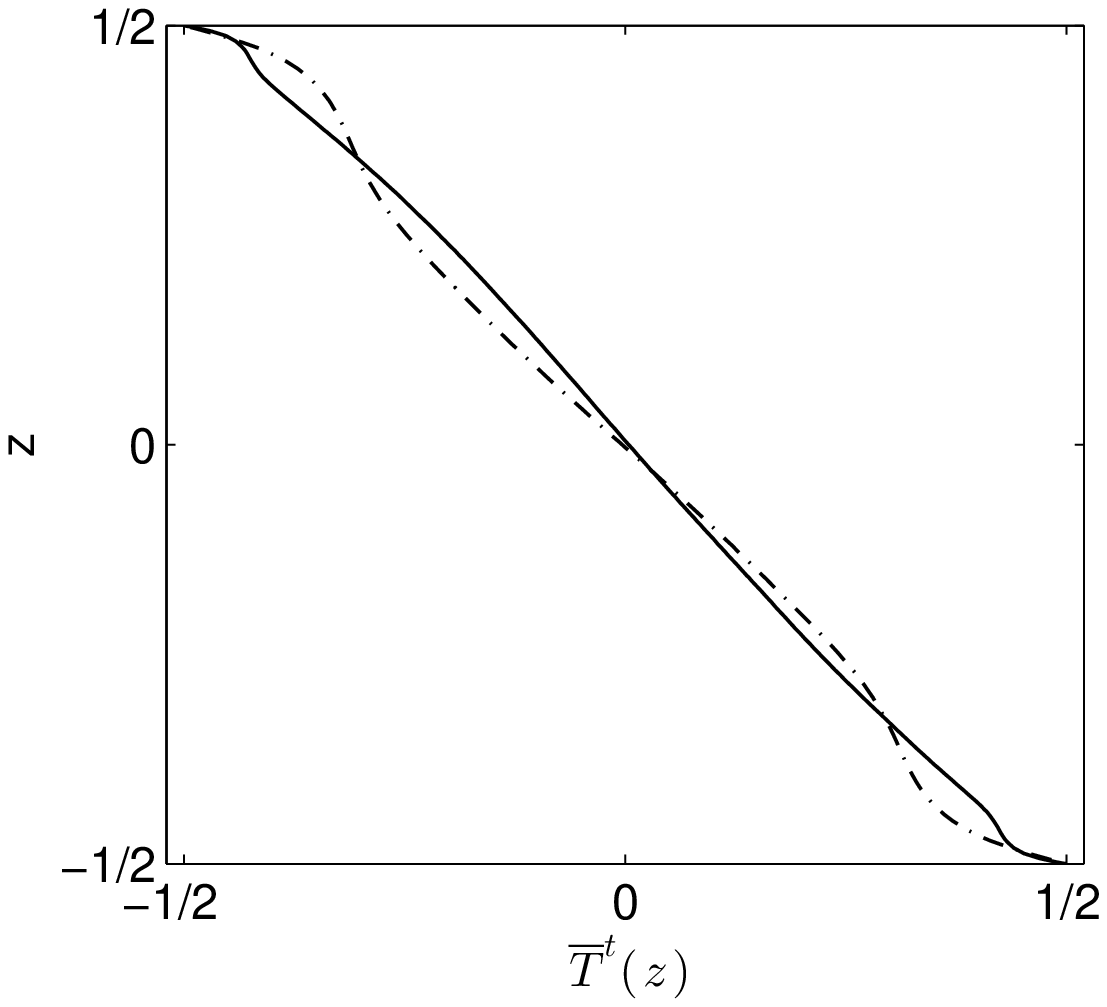}
\caption{Mean vertical profiles of zonal flow (left) and temperature (right) for bursting shearing convection with $(A,\Pran)=(2,1)$ at Rayleigh numbers of $2\e6$ (\protect\dashdotrule) and $2\e8$ (\solidrule). Each zonal flow profile is normalized by its maximum value, $\overline{u}^t_{max}$. Reversing $\overline{u}^t(z)$ yields the profiles that arise when symmetry breaks in the opposite way. Figure \ref{fig: Pr=10 profiles} shows analogous profiles for \emph{sustained} shearing convection.}
\label{fig: Pr=1 profiles}
\end{figure}

Thermal plumes are localized in \emph{time} in bursting shearing convection, whereas they are localized in \emph{space} in the sustained shearing convection of \S\ref{sec: sustained}. In the sustained case, the \href{http://www-personal.umich.edu/~goluskin/convection-driven-shear-videos}{videos} accompanying figure \ref{fig: video stills} illustrate that plumes exist at all times but have limited vertical extent, and we have identified this extent with the regions of positive temperature gradient, here called mixing layers, in the $\overline{T}^t(z)$ profiles of figure \hbox{\ref{fig: Pr=10 profiles}(b)}. In the bursting case, there are no mixing layers; figure \ref{fig: burst} and the accompanying \href{http://www-personal.umich.edu/~goluskin/convection-driven-shear-videos}{video} show that plumes penetrate the entire layer during bursts and are almost totally dispersed in between. Time-averaging over the bursts yields $\overline{T}^t(z)$ profiles that are fairly close to linear, as shown in figure \hbox{\ref{fig: Pr=1 profiles}(b)}, with the deviation from linearity near the boundaries being created almost entirely by the bursts. On the other hand, the normalized zonal flow profiles, shown for the bursting case in figure \hbox{\ref{fig: Pr=1 profiles}(a)}, are similar to their sustained counterparts in figure \hbox{\ref{fig: Pr=10 profiles}(a)}.

\section{Nusselt numbers}
\label{sec: nusselt}

\begin{figure}
\centering
\includegraphics[width=240pt]{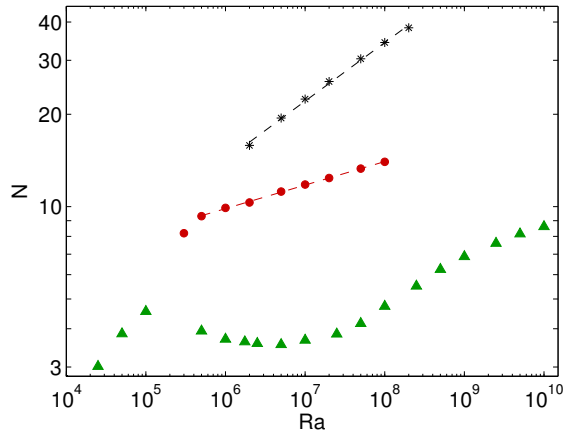}
\caption{Mean Nusselt numbers, $N$, of shearing convection at three Prandtl numbers and various Rayleigh numbers in a domain with $A=2$. The shearing convection is \emph{bursting} with $\Pran=1$ (\textcolor[rgb]{0,.6,0}{\scriptsize{$\blacktriangle$}}) and \emph{sustained} with $\Pran=3$ (\textcolor[rgb]{.8,0,0}{\scriptsize{$\bullet$}}) and $\Pran=10$ ($*$). In the sustained cases, dashed lines show power-law fits of $N\sim 3.4\,Ra^{0.077}$ for $\Pran=3$ and $N\sim 1.0\,Ra^{0.19}$ for $\Pran=10$. These $N$ values are compared to those of non-shearing states for $\Pran=10$ and $\Pran=1$ in figures \ref{fig: Pr=10 int} and \ref{fig: Pr=1 N}, respectively.}
\label{fig: N shear}
\end{figure}

Our findings on the mean Nusselt numbers of shearing convection are summarized in figure \ref{fig: N shear}, where the dependence of $N$ on $Ra$ is shown for Prandtl numbers of 1 (\textcolor[rgb]{0,.6,0}{\scriptsize{$\blacktriangle$}}), 3 (\textcolor[rgb]{.8,0,0}{\tiny{$\blacklozenge$}}), and 10 ($*$). The shearing convection is bursting when $\Pran=1$ and sustained when $\Pran=3$ or 10, as illustrated in figure \ref{fig: Ra=2e6} for the particular Rayleigh number of $2\e6$. At each $\Pran$, we found no shearing states at lower $Ra$ than shown in figure \ref{fig: N shear}.

The sharp change in the $Pr=1$ data of figure \ref{fig: N shear} seems to correspond to a transition from quasiperiodic states (in the three flows with lowest $Ra$) to chaotic bursting. We have found no quasiperiodic states with $Pr=10$, where shearing convection sets in at larger $Ra$. With $Pr$ smaller than unity, on the other hand, shearing convection sets in at smaller $Ra$ and can be quasiperiodic, periodic, or even steady \citep{Rucklidge1996, GoluskinThesis}. Such highly symmetric shearing states exist also at the parameters studied here but do not arise in simulations because they are unstable.

When $\Pran$ is smaller, the minimum Rayleigh number needed for shearing convection is also smaller. This is borne out not only by the $Ra$ at which shearing convection persists in direct numerical simulations (cf.\ figure \ref{fig: N shear}) but also by the $Ra$ at which \emph{steady} shearing states come into existence \citep{Rucklidge1996}. These steady states exist at the parameters studied here but do not arise in simulations because they are unstable. Steady shearing states that are stable have been found in our geometry only when $\Pran\lesssim0.3$ \citep{GoluskinThesis}. At somewhat larger $\Pran$, shearing states that are periodic or quasiperiodic in time can be stable. Of the states represented in figure \ref{fig: N shear}, it appears that the first with $\Pran=3$ and the first three with $\Pran=1$ are quasiperiodic, while the rest are chaotic. A more thorough study of these transitional states could be better carried out using the tools of numerical bifurcation analysis.

For all three Prandtl numbers represented in figure \ref{fig: N shear}, the dependence of $N$ on $Ra$ in shearing convection is unlike any reported for ordinary (non-shearing) RB convection. In the \emph{bursting} shearing convection with $\Pran=1$, $N$ varies \emph{non-monotonically} with $Ra$, decreasing in the range $10^5\lesssim Ra\lesssim2.5\e6$ before continuing to increase. This range of decrease is quite unusual, and we are not aware of a precedent. In the \emph{sustained} shearing convection, the Nusselt number grows roughly as a power of the Rayleigh number---our measured $N$ are best fit by power laws proportional to $Ra^{0.077}$ when $\Pran=3$ and to $Ra^{0.19}$ when $\Pran=10$. In the study of ordinary RB convection, many such power laws have been fit to measured $N$ and $Ra$, but the powers are always larger, ranging from $1/5$ to $1/2$ \citep{Grossmann2000, Ahlers2009}.

The only prior result on the scaling of $N(Ra,\Pran)$ in shearing convection, as far as we know, appears in the plasma-motivated work of \citet{Garcia2003a}, who simulated bursting shearing convection in a square domain with $\Pran=1$. Defining a suitable $N$ and $Ra$ for their thermal boundary conditions, we find that their $N$ grew monotonically (though slowly) with $Ra$, contrary to our findings at the same Prandtl number. As discussed in Appendix~\ref{app: Pr=1}, this difference might be caused by their narrower domain.

We have not found scaling arguments for $N(Ra,\Pran)$ in shearing convection that are consistent with the data of figure \ref{fig: N shear}. It may be easier to first study $N(Ra,\Pran)$ in \emph{non-shearing} RB convection between free-slip boundaries. Such convection has received much less attention than its no-slip counterpart, but the importance of velocity boundary conditions is hinted at by analytical upper bounds on $N$, whose scalings with $Ra$ can depend on the boundary conditions \citep{Otero2002a, Plasting2005, Ierley2006, Whitehead2011a, Whitehead2012}. In particular, $N$ cannot grow faster than $Ra^{5/12}$ when the boundaries are free-slip and the flow is two-dimensional \citep{Whitehead2011a}, so the flow cannot achieve the ultimate scaling of $Ra^{1/2}$ that is expected when the boundaries are no-slip \citep{Kraichnan1962, Spiegel1971b}. A viable alternative is the ultimate scaling of $Ra^{1/3}$ suggested by the argument, due to Priestly \citep[see][]{Spiegel1971a}, that $N$ should not depend on the layer thickness in the infinite-$Ra$ limit. At large but non-ultimate $Ra$, however, we are not aware of scaling arguments for RB convection between free-slip boundaries, be it shearing or non-shearing.

\section{Conclusions}
\label{sec: con}

Large-scale shear in RB convection is of interest not only because the RB model is an exemplar of complexity, but also because the interplay between convection and mean shear captures some aspects of more complicated astrophysical and plasma physical applications. We have studied the phenomenon by simulating a configuration in which it is especially pronounced; the free-slip boundaries and horizontal periodicity let mean horizontal flow travel unimpeded, and the two-dimensionality prevents disruption by motions transverse to the mean flow. When large-scale shear is present, we have referred to the total state as \emph{shearing convection} and to its mean horizontal component as the \emph{zonal flow}.

Although large-scale shear has been seen in laboratory convection \citep[for instance, by][]{Malkus1954, Krishnamurti1981}, the shear we have examined is different. Simulations with horizontal periodicity, which is impossible in cartesian laboratory experiments, allow for mean flows with horizontal wavenumbers of zero---the so-called zonal flows. These flows can account for a large fraction of the total kinetic energy and can significantly depress the vertical heat transport. Their name is adopted from the zonal flows that are seen in planetary atmospheres and tokamaks, and with which they share several features.

A direct comparison between shearing and non-shearing convection is possible in parameter regimes where states of both types are bistable, and such regimes exist for all Prandtl numbers between 1 and 10 that we have simulated in our chosen geometry. Shearing states are distinguished by the fact that plumes on opposing boundaries drift in opposite directions, as well as by the values of key integral quantities. Remarkably, the fraction of the total kinetic energy that the zonal flow accounts for tends towards unity as $Ra\to\infty$. The resulting large-scale shear helps disperse thermal plumes, thereby reducing heat transport. Nusselt numbers in shearing convection, compared to those of non-shearing convection, are smaller and grow more slowly with $Ra$, if they grow at all. Large-scale shear has been credited with depressing Nusselt numbers in a number of RB experiments \citep[such as those of][]{Castaing1989, Werne1993}, but the effect is more more pronounced here. It more resembles the dramatic reduction of heat transport that is possible when shear is imposed at the boundaries, rather than arising as a secondary flow \citep{Lipps1971, Doamaradzki1988, Zaleski1991}.

Shearing convection at large Rayleigh numbers exhibits one of two forms of time dependence. In what we call \emph{sustained} shearing convection, which occurs in our simulations when the Prandtl number is larger than about two, vertical convective transport is significant at all times. In \emph{bursting} shearing convection, which occurs at smaller Prandtl numbers, convective transport is confined to discrete bursts that resemble on-off intermittency \citep{Platt1993}. The bursting form of shearing convection has been described before, as surveyed by \citet{Garcia2006}, whereas it seems the sustained form has not. The zonal flow may be thought of as a parasite, draining energy from its convective host. In the sustained case, buoyancy forces are able to energize the host flow as quickly as the zonal flow parasitizes it. In the bursting case, the parasite drains energy too quickly, creating bursts as the host flow plummets and rebounds.

Several astrophysical questions may benefit from an improved understanding of shearing convection. For instance, helioseismic studies have revealed large-scale horizontal flows at the top and the bottom of the solar convection zone, and convection may play a role in driving them. The convective cores of massive stars rotate rapidly and so might also be poised to produce large-scale shear. If so, the shear would likely influence the lifetimes and internal structures of the stars since it has been found to augment reaction rates \citep{Spiegel1984}, much as it augments diffusion.

In tokamak plasmas, zonal flow is central to the effort to confine fusion magnetically \citep{Wagner2007}. Since tokamaks cannot yet be simulated under operating conditions, the construction of increasingly large experiments is justified, in part, by extrapolating past results. It is crucial that enlarging the tokamak not significantly enhance the interchange motions that harm confinement, at least in the ``high-confinement mode'' where zonal flow is present. The analogous behaviour in the RB system would be the Nusselt number of shearing convection holding steady as $Ra$ is raised. However, this is not what we observe---zonal flow slows, but often does not stop, the growth of the Nusselt number with $Ra$. This is especially true of the sustained form of shearing convection, which more resembles the high-confinement mode. [The bursting form of shearing convection has instead been likened to edge-localized modes in tokamaks \citep{Finn1993a, Leboeuf1993, Horton1996}.] The analogy between tokamaks and RB convection is only qualitative, and we cannot say how well it holds up. If taken at face value, our finding suggest that interchange motions would become increasingly harmful in ever-larger tokamaks, even in the high-confinement mode.

Open questions surround even basic features of large-scale shear in Rayleigh-B\'enard convection. Mapping out the parameter regimes in which shearing convection can be stable is not easy; the states in question are apparently chaotic when the domain is wide or the Prandtl number large, so transitions are hard to locate precisely, though our simulations give some estimates. Domains wider than ours have not been explored, and it is unclear how readily large-scale shear will persist in very wide domains. Once shearing convection is found in a region of parameter space, we can gain some insight into its character from integral quantities like the heat flux and the energy of the zonal flow. Beyond the data we have reported, little is known about the parameter-dependence of these quantities. An especially intriguing question is the fate of the Nusselt number in the infinite-$Ra$ limit. The ultimate scaling of $N\propto Ra^{1/2}$ conjectured for no-slip boundaries is ruled out for free-slip boundaries in two dimensions \citep{Whitehead2011a}, and even the unbounded growth of $N$ is not assured in shearing convection. Perhaps the real moral of our tale is that thermal convection continues to yield new aspects after centuries of study. It will no doubt remain a model of complexity for quite some time.

\vspace{10pt}
We are grateful for the interest and conversation of Charles Doering, Erwin van der Poel, Jared Whitehead, Keith Julien, and Edgar Knobloch, and for the suggestions of the anonymous referees. Da Zhu, in collaboration with G.R.F., performed many simulations with heat fluxes, rather than temperatures, fixed at the boundaries and found similar behaviour to what we have reported. The 2010 GFD summer program at the Woods Hole Oceanographic Institute was instrumental in our undertaking of this work. E.A.S. thanks Antonello Provenzale for some apposite remarks. D.G.\ is grateful for the general support of David Keyes and for the guidance of Paul Fischer and Aleksandr Obabko in running {\tt nek5000}. D.G.\ was supported during part of this work by NSF project EMSW21 -- RTG: Numerical Mathematics for Scientific Computing (Award No.\ DMS-0602235). Some computing resources were provided by the New York Center for Computational Sciences at Stony Brook University/Brookhaven National Laboratory, which is supported by the DOE (Contract No. DE-AC02-98CH10886) and by the State of New York. G.R.F.\ was supported by NSF collaborative proposal Models of the Deep Circulation of Gas Giants: Solar Heating, Convection, and Zonal Flows (Award No.\ AST-0708106).

\appendix

\section{Computational details}
\label{app: comp}

Simulations were carried out using both the \texttt{nek5000} spectral element code \citep{nek} and a fully spectral collocation code modified from that used by \citet{Johnston2009}. The two codes were verified against one another by simulating bursting shearing convection with $\Pran=1$ and $Ra\le10^7$. At each $Ra$, time-averaged Nusselt numbers agreed to at least two significant figures and usually to three. Bursting flows with $Ra>10^7$, which are the most numerically challenging flows studied here, were simulated using only the fully spectral code. Higher-$\Pran$ flows, which do not burst and are less numerically challenging, were simulated primarily using \texttt{nek5000} because the code scales well to large numbers of processors, though some cases were verified using the spectral code also.

The \texttt{nek5000} code was configured for second-order variable time stepping with a target Courant number of 1/2. Without bursting, time averages were deemed converged after a time span, $\tau$, when cumulative averages of the Nusselt number and kinetic energy, $N$ and $E$, agreed with their values after $\tau/2$ to within 0.2\%. Spatial meshes, composed of square and uniform elements, were deemed converged when increasing the polynomial order of each element by 2 changed $N$ and $E$ by less than 1\%. Most simulations were performed in a domain of aspect ratio $A=2$ with $64\times32$ elements. Elements of polynomial order 8 were needed when $Ra\ge10^8$, while elements of order 6 or 4 sufficed at smaller $Ra$. Table \ref{tab: Pr=10} gives representative parameters and Nusselt numbers for simulations of sustained shearing convection with $(A,\Pran)=(2,10)$.

\begin{table}
\begin{center}
\begin{tabular}{cccc}
$Ra$  	&$\tau$	& order & $N$ \\[3pt]
$2\e6$	& 3~~~	& 6		& 15.83 \\
$2\e7$	& 0.5~	& 6		& 25.59 \\
$2\e8$	& 0.18	& 8		& 38.28 \\
\end{tabular}
\caption{Representative parameters for simulations, using \texttt{nek5000}, of sustained shearing convection with $(A,\Pran)=(2,10)$. Quantities shown for each $Ra$ are the dimensionless runtime after transients ($\tau$), element order, and time-averaged $N$. All three simulations used a uniform mesh of $64\times32$ square elements. These three flows are used as examples throughout \S\ref{sec: sustained} and in figure~\ref{fig: Pr=10}.}
\label{tab: Pr=10}
\end{center}
\end{table}

The meshes needed to resolve sustained shearing convection are much coarser than demanded by non-shearing convection at similar $Ra$, and guidelines based on ordinary RB convection \citep{Shishkina2010} overestimate the resolution needed for our $N$ and $E$ to converge. This is reasonable since the shear leads to thicker thermal boundary layers, and it disperses fine structures. To further check mesh convergence in our \texttt{nek5000} simulations, the three highest-$Ra$ simulations were repeated using the fully spectral code, and $N$ was computed using four different expressions that should agree in the infinite-time limit. The results for $Ra=2\e8$ are shown in table \ref{tab: Pr=10 mesh}. Among both codes, both spectral meshes, and all four methods of calculating $N$, the greatest percent error was less than~1\%. 

\begin{table}
\begin{center}
\begin{tabular}{cccccccc}
code &	$n_x\times n_z$	& $\tau$	& BL points	& $1+\wT^t$		& 
$-\frac{d\oT^t}{dz}\Big|_B$	& ~$1+\frac{1}{Ra}\langle |\nabla\bu|^2\rangle^t$~ & $\langle |\nabla T|^2\rangle^t$ \\[3pt]
\texttt{nek5000}	& $64\times32$, order 8			& 0.18~\,	& 25 	& 38.28 \\
spectral 		& $512\times192$~~~~~~~~~~~	& 0.034	& 9		& 38.34	& 38.40& 38.41	& 38.41 \\
spectral 		& $768\times256$~~~~~~~~~~~	& 0.024	& 12	& 38.24	& 38.27	& 38.31	& 38.23
\end{tabular}
\caption{Verification of the Nusselt number in sustained shearing convection with $(A,\Pran,Ra)=(2,10,2\e8)$. Tabulated quantities include the mesh size ($n_x\times n_z$), dimensionless runtime after transients ($\tau$), the estimated number of points in the thermal boundary layer, and four expressions for $N$ that should be equal in the infinite-time limit.}
\label{tab: Pr=10 mesh}
\end{center}
\end{table}

The spectral collocation scheme we used to simulate the bursting shearing convection of \S\ref{sec: bursting} uses a Fourier basis in $x$ and a Chebyshev basis in $z$. The method is as described by \citet{Johnston2009}, except that we impose free-slip conditions here. Since the scheme integrates the vorticity-stream function ($\omega$-$\psi$) formulation of (\ref{eq: inc})-(\ref{eq: T}), impenetrable free-slip conditions are enforced easily by
\begin{equation}
\omega = \psi = 0 \text{ at } z=\pm\tfrac{1}{2}. \label{eq: free-slip stream}
\end{equation}
A tensor product of dimensions $n_x\times n_z$ was used for the spatial mesh. When $n_z\ge128$, the ordinary Chebyshev spacing in $z$ was modified by a coordinate mapping similar to that of \citet{Kosloff1993}. When $n_x\ge256$, an eighth order exponential filter was used to control aliasing errors in the convection term. Table \ref{tab: Pr=1} gives representative parameters and Nusselt numbers for simulations of bursting shearing convection with $(A,\Pran)=(2,1)$. Time averages were deemed converged after a time span, $\tau$, when $N$ and $E$ agreed with their values after $\tau/2$ to within 2\%. Spatial resolution was tested in a square domain ($A=1$) for $Ra\le10^8$, and meshes were deemed converged when increasing the resolution by $1/3$ changed $N$ by less than 1.5\%. Table \ref{tab: Pr=1 mesh} gives the results of these resolution tests.

\begin{table}
\begin{center}
\def~{\hphantom{0}}
\begin{tabular}{lccccccc}
$Ra$  		&  ~$n_x \times n_z$	& $\tau$& bursts	& $\Delta z_{min}$ & $N$  &  error \\[3pt]
$10^{6}$	&  ~192 $\times$ 96~	& 14.35 & 100 		&	5.78e-04	& 3.688 & 0.1\% \\ 
$10^{7}$	& ~384 $\times$ 160	& ~7.93 & 60   &	6.50e-04	& 3.670 & 0.1\%	 \\
$10^{8}$	& ~512 $\times$ 192	& ~2.43 & 50   &	2.56e-04	& 4.734 & 0.1\%	\\
$10^{9}$	& 1024 $\times$ 320	& ~0.73 & 45   &	1.46e-04	& 6.850 & 1.1\%	  \\
$10^{10}$	& 1536 $\times$ 448	& ~0.19 & 26   &	1.02e-04	& 8.485 & 1.8\%	  \\ 
\end{tabular}
\caption{Representative parameters for simulations, using the spectral collocation code, of bursting shearing convection with $(A,\Pran)=(2,1)$. Quantities shown for each $Ra$ are the mesh size ($n_x\times n_z$), dimensionless runtime after transients ($\tau$), number of bursts, minimum $z$-spacing of the mapped mesh ($\Delta z_{min}$), Nusselt number ($N$), and the relative error between $N$ after runtime $\tau/2$ and after runtime $\tau$. Values of $N$ computed using the temperature gradient at a boundary, instead of using $1+\wT^t$, differ by less than 1\%.}
\label{tab: Pr=1}
\end{center}
\end{table}

\begin{table}
\begin{center}
\begin{tabular}{ccccccc}
$Ra$  	& $n_x \times n_z$	& $\tau$	& bursts	& BL points	& $1+\wT^t$	& $-\frac{d\oT^t}{dz}\Big|_B$ \\[5pt]
$10^6$	& 128$\times$96~	& 9.6	& 69	& 10	& 2.707		& 2.701 \\
		& 192$\times$128	& 8.6	& 62	& 13	& 2.707		& 2.706 \\
$10^7$ & 256$\times$192	& 6.5	& 65	& 13	& 2.990		& 2.989 \\
		& 384$\times$256	& 1.9	& 19	& 18	& 2.995		& 2.992 \\
$10^8$	& 256$\times$192	& 2.6	& 64	& 11	& 4.360		& 4.359 \\
		& 512$\times$256	& 1.8	& 45	& 16	& 4.313		& 4.311  \\
\end{tabular}
\caption{Mesh convergence studies for simulations, using the spectral collocation code, of bursting shearing convection with $(A,\Pran)=(1,1)$. Quantities shown for each $Ra$ are the mesh size ($n_x\times n_z$), dimensionless runtime after transients ($\tau$), number of bursts, the estimated number of points in the thermal boundary layer during the peak of a larger-than-average burst, and two expressions for $N$ that should be equal in the infinite-time limit.}
\label{tab: Pr=1 mesh}
\end{center}
\end{table}

\section{Supplementary results}
\label{app: supp}

\subsection{Time series for sustained shearing convection with $\Pran=10$}
\label{app: time series}

\begin{figure}
\begin{tabular}{rrr}
$Ra=2\cdot10^6$: \hspace{23pt} &
$Ra=2\cdot10^7$: \hspace{23pt} &
$Ra=2\cdot10^8$:  \hspace{23pt}\vspace{5pt} \\
\includegraphics[width=122.5pt]{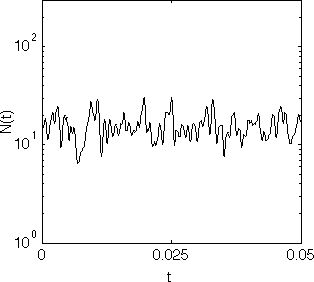} &
\includegraphics[width=122.5pt]{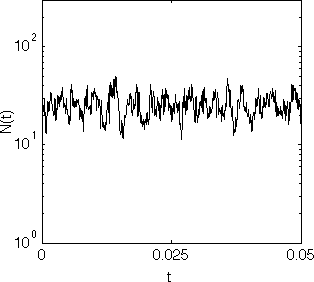} &
\includegraphics[width=122.5pt]{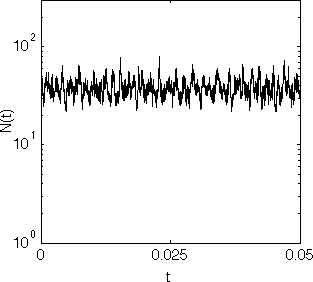} \vspace{2pt}\\
\includegraphics[width=115pt]{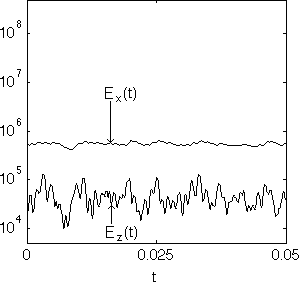} &
\includegraphics[width=115pt]{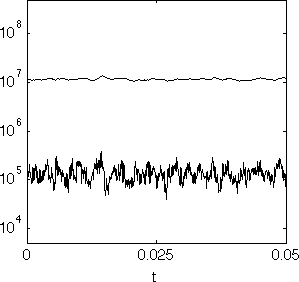} &
\includegraphics[width=115pt]{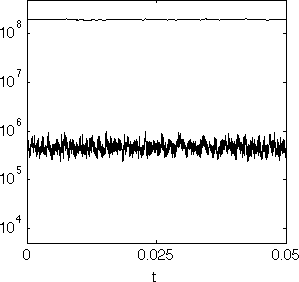}
\end{tabular}
\caption{Time series of the Nusselt numbers (top row) and the kinetic energies (bottom row) due to horizontal velocities ($E_x$, upper curves) and vertical velocities ($E_z$, lower curves) in sustained shearing convection with $(A,\Pran)=(2,10)$. The three Rayleigh numbers represented are $2\cdot10^6$ (left column), $2\cdot10^7$ (centre column), and $2\cdot10^8$ (right column). The integral quantities are as defined as in expressions (\ref{eq: E}) and (\ref{eq: N}) but with time averages not taken.}
\label{fig: Pr=10}
\end{figure}

Figure \ref{fig: Pr=10} shows time series of $N(t)$, $E_x(t)$, and $E_z(t)$ for the three examples of \emph{sustained} shearing convection whose temperature fields and time-averaged properties are explored in \S\ref{sec: sustained}. These series may be contrasted with the \emph{bursting} time series shown in figure \ref{fig: Pr=1}. Only fast local fluctuations are present in figure \ref{fig: Pr=10}, as opposed to bursts, and all three quantities fluctuate less strongly as $Ra$ is raised.

\subsection{Effects of boundary conditions and aspect ratio with $\Pran=1$}
\label{app: Pr=1}

\begin{figure}
\centering
\includegraphics[width=240pt]{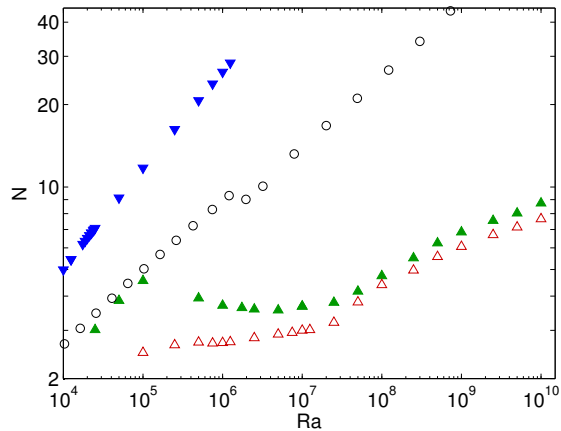}
\caption{Mean Nusselt numbers, $N$, for various Rayleigh numbers with $\Pran=1$. For $A=2$, bursting shearing convection (\textcolor[rgb]{0,.6,0}{\scriptsize{$\blacktriangle$}}) and non-shearing convection (\textcolor{blue}{\scriptsize{$\blacktriangledown$}}) are represented. For $A=1$, only bursting shearing convection (\textcolor[rgb]{.8,0,0}{\tiny$\triangle$}) is represented. Nusselt numbers computed by \citet{Johnston2009} for \emph{no-slip} boundaries and $A=2$ are also shown ({\large$\circ$}).}
\label{fig: Pr=1 N}
\end{figure}

Figure \ref{fig: Pr=1 N} shows Nusselt numbers of RB convection under several different conditions, all with $Pr=1$. The bursting shearing states shown in the figure (\textcolor[rgb]{0,.6,0}{\scriptsize{$\blacktriangle$}}) appear also in figure \ref{fig: N shear} alongside shearing states for other Prandtl numbers. Here, they are compared to non-shearing states (\textcolor{blue}{\scriptsize{$\blacktriangledown$}}) under the same conditions. Nusselt numbers are larger and grow much faster with $Ra$ in the non-shearing case, as is also true at other Prandtl numbers (cf.\ figure \ref{fig: Pr=10 int}). When the horizontal period is halved to $A=1$, the Nusselt numbers of shearing convection (\textcolor[rgb]{.8,0,0}{\tiny$\triangle$}) become monotonic in $Ra$.

The monotonic growth of $N$ that we find with $(A,\Pran)=(1,1)$ is consistent with the simulations of \citet{Garcia2003a}, where $A$ and $Pr$ were the same, but the heat flux, rather than the temperature, was fixed at the top boundary. To see this, we must reformulate the findings of \citet{Garcia2003a} in terms of $N$ and $Ra$. They report a fit equivalent to $\langle T\rangle^t\propto R^{-0.08}$, where $R$ is a Rayleigh number defined using the temperature difference between the boundaries in the static state. If we define $N$ and $Ra$ using the temperature difference between the boundaries in the developed convection, as is suited to comparing flows with different thermal boundary conditions \citep{Otero2002, Johnston2009, Wittenberg2010}, then the Nusselt number is the inverse of the temperature difference between the boundaries, and $Ra\propto R/N$. Since this temperature difference scales like $\langle T\rangle^t$ at large Rayleigh numbers, $N\propto1/\langle T\rangle^t$, and $Ra\propto R\langle T\rangle^t$. In these terms, the findings of \citeauthor{Garcia2003a} are that $N\propto Ra^{0.087}$. That is, $N$ increases slowly but monotonically with $Ra$, much as in our simulations.

When no-slip, rather than free-slip, conditions are imposed at the boundaries, only non-shearing states have been reported when $(A,\Pran)=(2,1)$. The Nusselt numbers of these non-shearing states ({\large$\circ$}) are smaller than those of non-shearing convection between free-slip boundaries (\textcolor{blue}{\scriptsize{$\blacktriangledown$}}) but larger than those of shearing convection between free-slip boundaries (\textcolor[rgb]{.8,0,0}{\tiny$\triangle$}).

\bibliographystyle{jfm}
\bibliography{library}

\begin{thebibliography}{74}
\expandafter\ifx\csname natexlab\endcsname\relax\def\natexlab#1{#1}\fi

\bibitem[Ahlers {\em et~al.\/}(2009)Ahlers, Grossmann \& Lohse]{Ahlers2009}
{\sc Ahlers, G., Grossmann, S. \& Lohse, D.} 2009 {Heat transfer and large
  scale dynamics in turbulent Rayleigh-B\'{e}nard convection}. {\em Rev. Mod.
  Phys.\/} {\bf 81}~(2), 503--537.

\bibitem[Bian {\em et~al.\/}(2003)Bian, Benkadda, Garcia, Paulsen \&
  Garbet]{Bian2003a}
{\sc Bian, N., Benkadda, S., Garcia, O.~E., Paulsen, J.-V. \& Garbet, X.} 2003
  {The quasilinear behavior of convective turbulence with sheared flows}. {\em
  Phys. Plasmas\/} {\bf 10}~(5), 1382--1388.

\bibitem[Bian \& Garcia(2003)]{Bian2003}
{\sc Bian, N.~H. \& Garcia, O.~E.} 2003 {Confinement and dynamical regulation
  in two-dimensional convective turbulence}. {\em Phys. Plasmas\/} {\bf
  10}~(12), 4696--4707.

\bibitem[Brummell \& Hart(1993)]{Brummell1993}
{\sc Brummell, N.~H. \& Hart, J.~E.} 1993 {High Rayleigh number
  $\beta$-convection}. {\em Geophys. Astrophys. Fluid Dyn.\/} {\bf 68},
  85--114.

\bibitem[Busse(1983)]{Busse1983}
{\sc Busse, F.~H.} 1983 {Generation of mean flows by thermal convection}. {\em
  Phys. D\/} {\bf 9}, 287--299.

\bibitem[Busse(1994)]{Busse1994}
{\sc Busse, F.~H.} 1994 {Convection driven zonal flows and vortices in the
  major planets}. {\em Chaos\/} {\bf 4}~(2), 123--134.

\bibitem[Castaing {\em et~al.\/}(1989)Castaing, Gunaratne, Heslot, Kadanoff,
  Libchaber, Thomae, Wu, Zaleski \& Zanetti]{Castaing1989}
{\sc Castaing, B., Gunaratne, G., Heslot, F., Kadanoff, L., Libchaber, A.,
  Thomae, S., Wu, X.-Z., Zaleski, S. \& Zanetti, G.} 1989 {Scaling of hard
  thermal turbulence in Rayleigh-B\'{e}nard convection}. {\em J. Fluid Mech.\/}
  {\bf 204}~(1), 1--30.

\bibitem[Chaboyer \& Zahn(1992)]{Chaboyer1992}
{\sc Chaboyer, B. \& Zahn, J.-P.} 1992 {Effect of horizontal turbulent
  diffusion on transport by meridional circulation}. {\em Astron. Astrophys.\/}
  {\bf 253}, 173--177.

\bibitem[Chandrasekhar(1981)]{Chandrasekhar1981}
{\sc Chandrasekhar, S.} 1981 {\em {Hydrodynamic and Hydromagnetic
  Stability}\/}. Dover Publications.

\bibitem[Childress(2000)]{Childress2000}
{\sc Childress, S.} 2000 {Eulerian mean flow from an instability of convective
  plumes.} {\em Chaos\/} {\bf 10}~(1), 28--38.

\bibitem[Cho \& Polvani(1996)]{Polvani1996}
{\sc Cho, J. Y.-K. \& Polvani, L.~M.} 1996 {The emergence of jets and vortices
  in freely evolving, shallow-water turbulence on a sphere}. {\em Phys.
  Fluids\/} {\bf 8}~(6), 1531--1552.

\bibitem[Christensen(2002)]{Christensen2002}
{\sc Christensen, U.~R.} 2002 {Zonal flow driven by strongly supercritical
  convection in rotating spherical shells}. {\em J. Fluid Mech.\/} {\bf 470},
  115--133.

\bibitem[Deardorff(1965)]{Deardorff1965}
{\sc Deardorff, J.~W.} 1965 {Graviational instability between horizontal plates
  with shear}. {\em Phys. Fluids\/} {\bf 8}~(6), 1027--1030.

\bibitem[Diamond {\em et~al.\/}(2005)Diamond, Itoh, Itoh \& Hahm]{Diamond2005}
{\sc Diamond, P.~H., Itoh, S.-I., Itoh, K. \& Hahm, T.~S.} 2005 {Zonal flows in
  plasma---a review}. {\em Plasma Phys. Control. Fusion\/} {\bf 47}~(5),
  R35--R161.

\bibitem[Domaradzki(1988)]{Doamaradzki1988}
{\sc Domaradzki, J.} 1988 {Direct numerical simulations of the effects of shear
  on turbulent Rayleigh-B\'{e}nard convection}. {\em J. Fluid Mech.\/} {\bf
  193}, 499--531.

\bibitem[Drake {\em et~al.\/}(1992)Drake, Finn, Guzdar, Shapiro, Shevchenko,
  Waelbroeck, Hassam, Liu \& Sagdeev]{Drake1992}
{\sc Drake, J.~F., Finn, J.~M., Guzdar, P., Shapiro, V., Shevchenko, V.,
  Waelbroeck, F., Hassam, A.~B., Liu, C.~S. \& Sagdeev, R.} 1992 {Peeling of
  convection cells and the generation of sheared flow}. {\em Phys. Fluids B
  Plasma Phys.\/} {\bf 4}~(3), 488--491.

\bibitem[Finn(1993)]{Finn1993a}
{\sc Finn, J.~M.} 1993 {Nonlinear interaction of Rayleigh-Taylor and shear
  instabilities}. {\em Phys. Fluids B Plasma Phys.\/} {\bf 5}~(2), 415--432.

\bibitem[Finn {\em et~al.\/}(1992)Finn, Drake \& Guzdar]{Finn1992}
{\sc Finn, J.~M., Drake, J.~F. \& Guzdar, P.~N.} 1992 {Instability of fluid
  vortices and generation of sheared flow}. {\em Phys. Fluids B Plasma Phys.\/}
  {\bf 4}~(9), 2758--2768.

\bibitem[Fisher {\em et~al.\/}(2014)Fisher, Lottes \& Kerkemeier]{nek}
{\sc Fisher, P.~F., Lottes, J.~W. \& Kerkemeier, S.~G.} 2014 {{\tt nek5000} Web
  page}.

\bibitem[Fitzgerald \& Farrell(2014)]{Fitzgerald2014}
{\sc Fitzgerald, J.~G. \& Farrell, B.~F.} 2014 {Mechanisms of mean flow
  formation and suppression in two-dimensional Rayleigh-B\'{e}nard convection}.
  {\em Phys. Fluids\/} {\bf 26}~(5), 054104.

\bibitem[Gallagher \& Mercer(1965)]{Gallagher1965}
{\sc Gallagher, A.~P. \& Mercer, A.~McD.} 1965 {On the behaviour of small
  disturbances in plane Couette flow with a temperature gradient}. {\em Proc.
  R. Soc. London Ser. A\/} {\bf 286}~(1404), 117--128.

\bibitem[Garcia \& Bian(2003)]{Garcia2003}
{\sc Garcia, O.~E. \& Bian, N.~H.} 2003 {Bursting and large-scale intermittency
  in turbulent convection with differential rotation}. {\em Phys. Rev. E\/}
  {\bf 68}~(4), 1--4.

\bibitem[Garcia {\em et~al.\/}(2006)Garcia, Bian, Naulin, Nielsen \&
  Rasmussen]{Garcia2006}
{\sc Garcia, O.~E., Bian, N.~H., Naulin, V., Nielsen, A.~H. \& Rasmussen,
  J.~J.} 2006 {Two-dimensional convection and interchange motions in fluids and
  magnetized plasmas}. {\em Phys. Scr.\/} {\bf T122}, 104--124.

\bibitem[Garcia {\em et~al.\/}(2003)Garcia, Bian, Paulsen, Benkadda \&
  Rypdal]{Garcia2003a}
{\sc Garcia, O.~E., Bian, N.~H., Paulsen, J.-V., Benkadda, S. \& Rypdal, K.}
  2003 {Confinement and bursty transport in a flux-driven convection model with
  sheared flows}. {\em Plasma Phys. Control. Fusion\/} {\bf 45}, 919--932.

\bibitem[Goluskin(2013)]{GoluskinThesis}
{\sc Goluskin, D.} 2013 {Zonal flow driven by convection and convection driven
  by internal heating}. PhD thesis, Columbia University.

\bibitem[Grossmann \& Lohse(2000)]{Grossmann2000}
{\sc Grossmann, S. \& Lohse, D.} 2000 {Scaling in thermal convection: a
  unifying theory}. {\em J. Fluid Mech.\/} {\bf 407}, 27--56.

\bibitem[von Hardenberg {\em et~al.\/}(2008)von Hardenberg, Parodi, Passoni,
  Provenzale \& Spiegel]{VonHardenberg2008}
{\sc von Hardenberg, J., Parodi, A., Passoni, G., Provenzale, A. \& Spiegel,
  E.~A.} 2008 {Large-scale patterns in Rayleigh-B\'{e}nard convection}. {\em
  Phys. Lett. A\/} {\bf 372}~(13), 2223--2229.

\bibitem[Heimpel \& Aurnou(2007)]{Heimpel2007}
{\sc Heimpel, M. \& Aurnou, J.} 2007 {Turbulent convection in rapidly rotating
  spherical shells: a model for equatorial and high latitude jets on Jupiter
  and Saturn}. {\em Icarus\/} {\bf 187}, 540--557.

\bibitem[Hermiz {\em et~al.\/}(1995)Hermiz, Guzdar \& Finn]{Hermiz1995}
{\sc Hermiz, K.~B., Guzdar, P.~N. \& Finn, J.~M.} 1995 {Improved low-order
  model for shear flow driven by Rayleigh-B\'{e}nard convection}. {\em Phys.
  Rev. E\/} {\bf 51}~(1), 325--331.

\bibitem[Horton {\em et~al.\/}(1996)Horton, Hu \& Laval]{Horton1996}
{\sc Horton, W., Hu, G. \& Laval, G.} 1996 {Turbulent transport in mixed states
  of convective cells and sheared flows}. {\em Phys. Plasmas\/} {\bf 3}~(8),
  2912--2923.

\bibitem[Howard(1963)]{Howard1963}
{\sc Howard, L.~N.} 1963 {Heat transport by turbulent convection}. {\em J.
  Fluid Mech.\/} {\bf 17}~(3), 405--432.

\bibitem[Howard \& Krishnamurti(1986)]{Howard1986}
{\sc Howard, L.~N. \& Krishnamurti, R.} 1986 {Large-scale flow in turbulent
  convection: a mathematical model}. {\em J. Fluid Mech.\/} {\bf 170}~(1),
  385--410.

\bibitem[Ierley {\em et~al.\/}(2006)Ierley, Kerswell \& Plasting]{Ierley2006}
{\sc Ierley, G.~R., Kerswell, R.~R. \& Plasting, S.~C.} 2006
  {Infinite-Prandtl-number convection. part 2. a singular limit of upper bound
  theory}. {\em J. Fluid Mech.\/} {\bf 560}, 159--227.

\bibitem[Ingersoll(1966)]{Ingersoll1966}
{\sc Ingersoll, A.~P.} 1966 {Convective instabilities in plane Couette flow}.
  {\em Phys. Fluids\/} {\bf 9}~(4), 682--689.

\bibitem[Johnston \& Doering(2009)]{Johnston2009}
{\sc Johnston, H. \& Doering, C.~R.} 2009 {Comparison of turbulent thermal
  convection between conditions of constant temperature and constant flux}.
  {\em Phys. Rev. Lett.\/} {\bf 102}~(6), 064501.

\bibitem[Kaspi {\em et~al.\/}(2009)Kaspi, Flierl \& Showman]{Kaspi2009}
{\sc Kaspi, Y., Flierl, G.~R. \& Showman, A.~P.} 2009 {The deep wind structure
  of the giant planets: results from an anelastic general circulation model}.
  {\em Icarus\/} {\bf 202}~(2), 525--542.

\bibitem[Knobloch \& Moehlis(1999)]{Knobloch1999a}
{\sc Knobloch, E. \& Moehlis, J.} 1999 {Bursting mechanisms for hydrodynamical
  systems}. In {\em Pattern Formation in Continuous and Coupled Systems\/}, pp.
  157--174. Springer.

\bibitem[Kosloff \& Tal-Ezer(1993)]{Kosloff1993}
{\sc Kosloff, D. \& Tal-Ezer, H.} 1993 {A modified Chebyshev pseudospectral
  method with an $O(N^{-1})$ time step restriction}. {\em J. Comput. Phys.\/}
  {\bf 104}~(2), 457--469.

\bibitem[Kraichnan(1962)]{Kraichnan1962}
{\sc Kraichnan, R.~H.} 1962 {Turbulent thermal convection at arbitrary Prandtl
  number}. {\em Phys. Fluids\/} {\bf 5}~(11), 1374--1389.

\bibitem[Krishnamurti \& Howard(1981)]{Krishnamurti1981}
{\sc Krishnamurti, R. \& Howard, L.~N.} 1981 {Large-scale flow generation in
  turbulent convection}. {\em Proc. Natl. Acad. Sci. U. S. A.\/} {\bf 78}~(4),
  1981--1985.

\bibitem[Leboeuf {\em et~al.\/}(1993)Leboeuf, Charlton \&
  Carreras]{Leboeuf1993}
{\sc Leboeuf, J.-N., Charlton, L.~A. \& Carreras, B.~A.} 1993 {Shear flow
  effects on the nonlinear evolution of thermal instabilities}. {\em Phys.
  Fluids B Plasma Phys.\/} {\bf 5}~(8), 2959--2966.

\bibitem[Lipps(1971)]{Lipps1971}
{\sc Lipps, F.~B.} 1971 {Two-dimensional numerical experiments in thermal
  convection with vertical shear}. {\em J. Atmos. Sci.\/} {\bf 28}~(1), 3--19.

\bibitem[Lotka(1925)]{Lotka1925}
{\sc Lotka, A.~J.} 1925 {\em {Elements of Physical Biology}\/}. Williams and
  Wilkins Co.

\bibitem[Malkov {\em et~al.\/}(2001)Malkov, Diamond \& Rosenbluth]{Malkov2001}
{\sc Malkov, M.~A., Diamond, P.~H. \& Rosenbluth, M.~N.} 2001 {On the nature of
  bursting in transport and turbulence in drift wave–zonal flow systems}.
  {\em Phys. Plasmas\/} {\bf 8}~(12), 5073--5076.

\bibitem[Malkus(1954{\natexlab{{\em a\/}}})]{Malkus1954}
{\sc Malkus, W. V.~R.} 1954{\natexlab{{\em a\/}}} {Discrete transitions in
  turbulent convection}. {\em Proc. R. Soc. London Ser. A\/} {\bf 225}~(1161),
  185--195.

\bibitem[Malkus(1954{\natexlab{{\em b\/}}})]{Malkus1954a}
{\sc Malkus, W. V.~R.} 1954{\natexlab{{\em b\/}}} {The heat transport and
  spectrum of thermal turbulence}. {\em Proc. R. Soc. London Ser. A\/} {\bf
  225}~(1161), 196--212.

\bibitem[Massaguer {\em et~al.\/}(1992)Massaguer, Spiegel \&
  Zahn]{Massaguer1992}
{\sc Massaguer, J.~M., Spiegel, E.~A. \& Zahn, J.-P.} 1992 {Convection-induced
  shears for general planforms}. {\em Phys. Fluids A\/} {\bf 4}~(7),
  1333--1335.

\bibitem[Matthews {\em et~al.\/}(1996)Matthews, Rucklidge, Weiss \&
  Proctor]{Matthews1996}
{\sc Matthews, P.~C., Rucklidge, A.~M., Weiss, N.~O. \& Proctor, M. R.~E.} 1996
  {The three-dimensional development of the shearing instability of
  convection}. {\em Phys. Fluids\/} {\bf 8}~(6), 1350--1352.

\bibitem[Morin \& Dormy(2004)]{Morin2004}
{\sc Morin, V. \& Dormy, E.} 2004 {Time dependent $\beta$-convection in rapidly
  rotating spherical shells}. {\em Phys. Fluids\/} {\bf 16}~(5), 1603--1609.

\bibitem[Otero(2002)]{Otero2002a}
{\sc Otero, J.} 2002 {Bounds for the heat transport in turbulent convection}.
  PhD thesis, University of Michigan.

\bibitem[Otero {\em et~al.\/}(2002)Otero, Wittenberg, Worthing \&
  Doering]{Otero2002}
{\sc Otero, J., Wittenberg, R.~W., Worthing, R.~A. \& Doering, C.~R.} 2002
  {Bounds on Rayleigh-B\'{e}nard convection with an imposed heat flux}. {\em J.
  Fluid Mech.\/} {\bf 473}, 191--199.

\bibitem[Parodi {\em et~al.\/}(2004)Parodi, von Hardenberg, Passoni, Provenzale
  \& Spiegel]{Parodi2004}
{\sc Parodi, A., von Hardenberg, J., Passoni, G., Provenzale, A. \& Spiegel,
  E.~A.} 2004 {Clustering of plumes in turbulent convection}. {\em Phys. Rev.
  Lett.\/} {\bf 92}~(19), 194503.

\bibitem[Plasting \& Ierley(2005)]{Plasting2005}
{\sc Plasting, S.~C. \& Ierley, G.~R.} 2005 {Infinite-Prandtl-number
  convection. part 1. conservative bounds}. {\em J. Fluid Mech.\/} {\bf 542},
  343--363.

\bibitem[Platt {\em et~al.\/}(1993)Platt, Spiegel \& Tresser]{Platt1993}
{\sc Platt, N., Spiegel, E.~A. \& Tresser, C.} 1993 {On-off intermittency: a
  mechanism for bursting}. {\em Phys. Rev. Lett.\/} {\bf 70}~(3), 279--282.

\bibitem[van~der Poel {\em et~al.\/}(2014)van~der Poel, Ostilla-M\'{o}nico,
  Verzicco \& Lohse]{VanderPoel2014a}
{\sc van~der Poel, E.~P., Ostilla-M\'{o}nico, R., Verzicco, R. \& Lohse, D.}
  2014 {Effect of velocity boundary conditions on the heat transfer and flow
  topology in two-dimensional Rayleigh-B\'{e}nard convection}. {\em Phys. Rev.
  E\/} {\bf 90}~(1), 013017.

\bibitem[Rucklidge \& Matthews(1996)]{Rucklidge1996}
{\sc Rucklidge, A.~M. \& Matthews, P.~C.} 1996 {Analysis of the shearing
  instability in nonlinear convection and magnetoconvection}. {\em
  Nonlinearity\/} {\bf 9}~(311-351).

\bibitem[Scagliarini {\em et~al.\/}(2014)Scagliarini, Gylfason \&
  Toschi]{Scagliarini2014}
{\sc Scagliarini, A., Gylfason, \'{A}. \& Toschi, F.} 2014 {Heat-flux scaling
  in turbulent Rayleigh-B\'{e}nard convection with an imposed longitudinal
  wind}. {\em Phys. Rev. E\/} {\bf 89}~(4), 043012.

\bibitem[Shishkina {\em et~al.\/}(2010)Shishkina, Stevens, Grossmann \&
  Lohse]{Shishkina2010}
{\sc Shishkina, O., Stevens, R. J. A.~M., Grossmann, S. \& Lohse, D.} 2010
  {Boundary layer structure in turbulent thermal convection and its
  consequences for the required numerical resolution}. {\em New J. Phys.\/}
  {\bf 12}~(7), 075022.

\bibitem[Spiegel(1971{\natexlab{{\em a\/}}})]{Spiegel1971a}
{\sc Spiegel, E.~A.} 1971{\natexlab{{\em a\/}}} {Convection in stars I. Basic
  Boussinesq convection}. {\em Annu. Rev. Astron. Astrophys.\/} {\bf 9},
  323--352.

\bibitem[Spiegel(1971{\natexlab{{\em b\/}}})]{Spiegel1971b}
{\sc Spiegel, E.~A.} 1971{\natexlab{{\em b\/}}} {Turbulence in stellar
  convection zones}. {\em Comments Astrophys. Space Phys.\/} {\bf 3}~(2).

\bibitem[Spiegel \& Zaleski(1984)]{Spiegel1984}
{\sc Spiegel, E.~A. \& Zaleski, S.} 1984 {Reaction-diffusion instability in a
  sheared medium}. {\em Phys. Lett. A\/} {\bf 106}~(7), 335--338.

\bibitem[Stevens {\em et~al.\/}(2013)Stevens, van~der Poel, Grossmann \&
  Lohse]{Stevens2013}
{\sc Stevens, R. J. A.~M., van~der Poel, E.~P., Grossmann, S. \& Lohse, D.}
  2013 {The unifying theory of scaling in thermal convection: the updated
  prefactors}. {\em J. Fluid Mech.\/} {\bf 730}, 295--308.

\bibitem[Tao \& Tan(2010)]{Tao2010}
{\sc Tao, J.-J. \& Tan, W.-C.} 2010 {Relaxation oscillation of thermal
  convection in rotating cylindrical annulus}. {\em Chinese Phys. Lett.\/} {\bf
  27}~(3), 034706.

\bibitem[Taylor(1953)]{Taylor1953}
{\sc Taylor, G.} 1953 {Dispersion of soluble matter in solvent flowing slowly
  through a tube}. {\em Proc. R. Soc. London Ser. A\/} {\bf 219}~(1137),
  186--203.

\bibitem[Terry(2000)]{Terry2000}
{\sc Terry, P.~W.} 2000 {Suppression of turbulence and transport by sheared
  flow}. {\em Rev. Mod. Phys.\/} {\bf 72}~(1), 109--165.

\bibitem[Thompson(1970)]{Thompson1970}
{\sc Thompson, R.} 1970 {Venus's general circulation is a merry-go-round}. {\em
  J. Atmos. Sci.\/} {\bf 27}~(8), 1107--1116.

\bibitem[Wagner(2007)]{Wagner2007}
{\sc Wagner, F.} 2007 {A quarter-century of H-mode studies}. {\em Plasma Phys.
  Control. Fusion\/} {\bf 49}~(12B), B1--B33.

\bibitem[Werne(1993)]{Werne1993}
{\sc Werne, J.} 1993 {Structure of hard-turbulent convection in two dimensions:
  numerical evidence}. {\em Phys. Rev. E\/} {\bf 48}~(2), 1020--1035.

\bibitem[Wesson(2011)]{Wesson2011}
{\sc Wesson, J.} 2011 {\em {Tokamaks}\/}, 4th edn. Oxford University Press.

\bibitem[Whitehead \& Doering(2011)]{Whitehead2011a}
{\sc Whitehead, J.~P. \& Doering, C.~R.} 2011 {Ultimate state of
  two-dimensional Rayleigh-B\'{e}nard convection between free-slip
  fixed-temperature boundaries}. {\em Phys. Rev. Lett.\/} {\bf 106}~(24),
  244501.

\bibitem[Whitehead \& Doering(2012)]{Whitehead2012}
{\sc Whitehead, J.~P. \& Doering, C.~R.} 2012 {Rigid bounds on heat transport
  by a fluid between slippery boundaries}. {\em J. Fluid Mech.\/} {\bf 707},
  241--259.

\bibitem[Wittenberg(2010)]{Wittenberg2010}
{\sc Wittenberg, R.~W.} 2010 {Bounds on Rayleigh-B\'{e}nard convection with
  imperfectly conducting plates}. {\em J. Fluid Mech.\/} {\bf 665}, 158--198.

\bibitem[Zaleski(1991)]{Zaleski1991}
{\sc Zaleski, S.} 1991 {Thermal convection at high Rayleigh numbers in two
  dimensional sheared layers}. In {\em The global geometry of turbulence\/}
  (ed. J.~Jim\'{e}nez), pp. 167--179. Springer.

\bibitem[Zhu \& Flierl(2012)]{Zhu2012}
{\sc Zhu, D. \& Flierl, G.~R.} 2012 {Investigation of vertically sheared flow
  in fixed-flux turbulent convection. Unpublished manuscript.}

\end{thebibliography}
\end{document}